\documentclass[twocolumn,superscriptaddress,amsmath, amssymb, amsfonts,preprintnumbers,aps,prd,longbibliography,nofootinbib]{revtex4-2}

\renewcommand{\sec}[1]{\textit{#1. --- }}

\usepackage{graphicx}
\usepackage{dcolumn}
\usepackage{bm}
\usepackage{amstext}
\usepackage{amssymb}
\usepackage[justification=centering]{subcaption}
\usepackage{amsmath}
\usepackage{graphicx}
\usepackage{color}
\usepackage{bbold}
\usepackage{delimset} 
\usepackage[colorlinks=true]{hyperref}

\usepackage{orcidlink}

\newcommand{\gusNotes}[1]{}

\DeclareFontFamily{OT1}{pzc}{}
\DeclareFontShape{OT1}{pzc}{m}{it}{<-> s * [1.350] pzcmi7t}{}
\DeclareMathAlphabet{\mathpzc}{OT1}{pzc}{m}{it}

\def\cN{\mathcal{N}}
\def\cO{\mathcal{O}}
\def\cD{\mathcal{D}}

\def\cE{\mathcal{E}}

\def\eps{\epsilon}

\def\d{\mathrm{d}}
\def\D{\mathrm{D}}

\def\dd{\delta\!\!\!{}^-\!}

\def\d{\mathrm{d}}
\def\eps{\epsilon}

\def\braket#1{\langle #1 \rangle}

\def\nn{\nonumber}

\def\eg{e.g. }

\def\Eqn#1{Eq.~\eqref{#1}}
\def\Eqns#1#2{Eqs.~\eqref{#1} and~\eqref{#2}}

\def\Fig#1{Fig.~{\ref{#1}}}

\def\Sec#1{Section~{\ref{#1}}}

\def\Rcite#1{Ref.~\cite{#1}}
\def\Rcites#1{Refs.~\cite{#1}}

\newcommand{\vct}[1]{\mathbf{#1}}

\newcommand{\be}{\begin{equation}}
\newcommand{\ee}{\end{equation}}
\newcommand{\ba}{\begin{align}}
\newcommand{\ea}{\end{align}}
\ifx\genfrac\sdflkaj\else\fi
\newcommand{\sfrac}[2]{{\textstyle\frac{#1}{#2}}}

\newcommand{\pin}{p_\infty}

\newcommand{\thii}[1]{\theta^{(#1)}}
\newcommand{\thiic}[1]{\theta^{(#1)}_{\rm can}}
\newcommand{\thiio}[1]{\theta^{(#1)}_{\rm cons}}
\newcommand{\thiir}[1]{\theta^{(#1)}_{\rm rad}}
\newcommand{\phiic}[1]{\phi^{(#1)}_{\rm can}}

\newcommand{\cii}[1]{c^{(#1)}}
\newcommand{\pc}{\phi_{\rm cons}}
\newcommand{\tc}{\theta_{\rm cons}}

\begin{document}

\preprint{HU-EP-22/32-RTG}

\title{Linear Response, Hamiltonian and Radiative Spinning Two-Body Dynamics}

\author{Gustav Uhre Jakobsen\,\orcidlink{0000-0001-9743-0442}}
\email{gustav.uhre.jakobsen@physik.hu-berlin.de}
\affiliation{%
  Institut f\"ur Physik und IRIS Adlershof, Humboldt Universit\"at zu Berlin,
  Zum Gro{\ss}en Windkanal 2, 12489 Berlin, Germany
}
\affiliation{Max Planck Institute for Gravitational Physics (Albert Einstein Institute), Am M\"uhlenberg 1, 14476 Potsdam, Germany}
\affiliation{Kavli Institute for Theoretical Physics, University of California, Santa Barbara, CA 93106, USA}

\author{Gustav Mogull\,\orcidlink{0000-0003-3070-5717}}
\email{gustav.mogull@aei.mpg.de} 
\affiliation{%
  Institut f\"ur Physik und IRIS Adlershof, Humboldt Universit\"at zu Berlin,
  Zum Gro{\ss}en Windkanal 2, 12489 Berlin, Germany
}
\affiliation{Max Planck Institute for Gravitational Physics (Albert Einstein Institute), Am M\"uhlenberg 1, 14476 Potsdam, Germany}
\affiliation{Kavli Institute for Theoretical Physics, University of California, Santa Barbara, CA 93106, USA}

\begin{abstract}
  Using the spinning, supersymmetric Worldline Quantum Field Theory formalism
  we compute the momentum impulse and spin kick from a scattering of
  two spinning black holes or neutron stars up to quadratic order in spin
  at third post-Minkowskian (PM) order, including radiation-reaction effects and with arbitrarily mis-aligned spin directions.
  Parts of these observables, both conservative and radiative,
  are also inferred from lower-PM scattering data by extending Bini and Damour's linear response formula
  to include mis-aligned spins.
  By solving Hamilton's equations of motion we also use a conservative scattering angle
  to infer a complete 3PM two-body Hamiltonian including finite-size corrections and mis-aligned spin-spin interactions.
  Finally, we describe mappings to the bound two-body dynamics for aligned spin vectors:
  including a numerical plot of the binding energy for circular orbits compared with numerical relativity,
  analytic confirmation of the NNLO PN binding energy and the energy loss over successive orbits.
\end{abstract}

\maketitle

The need for accurate waveform templates for comparison with gravitational wave
signals coming from the LIGO, Virgo and KAGRA detectors of binary merger events
\cite{Abbott:2016blz,LIGOScientific:2017vwq,LIGOScientific:2018mvr,LIGOScientific:2020ibl,LIGOScientific:2021usb,LIGOScientific:2021djp} ---
and in the future LISA, the Einstein Telescope and Cosmic Explorer~\cite{Ballmer:2022uxx} ---
has provoked enormous interest in the gravitational two-body problem.
One of the most important physical properties influencing the paths
of massive objects following inspiral trajectories,
which as they accelerate produce gravitational waves, is their spins.
Accurately determining the spins of black holes and neutron stars 
in binary orbits yields crucial information about their origins:
if the spins are approximately aligned with the orbital plane,
then this suggests formation of the binary system by slow accretion of matter;
if they are mis-aligned (precessing), then this indicates formation of the binary
by a random capture event.

A fruitful path has been effective field theory (EFT)-based methods,
which tackle the inspiral stage of the gravitational two-body problem using its natural separation of length scales
\cite{Goldberger:2004jt,Goldberger:2006bd,Goldberger:2009qd,
  Porto:2016pyg,Levi:2018nxp}:
the size of the massive bodies is far less than their separation,
which in turn is far less than their distance from us, the observer.
Partial results for the non-spinning two-body Hamiltonian are available up to sixth post-Newtonian (PN) order
\cite{Blumlein:2020znm,Bini:2020nsb,Bini:2020hmy,Blumlein:2020pyo,Foffa:2020nqe,Blumlein:2021txj};
in the spinning case a body-fixed frame on the worldline is often used~\cite{Porto:2005ac,Levi:2015msa,Goldberger:2020fot,Saketh:2022wap},
and results are available up to N$^3$LO in the spin-orbit sector~\cite{Levi:2020kvb,Kim:2022pou,Mandal:2022nty}
and in the spin-spin sector~\cite{Levi:2011eq,Levi:2014sba,Levi:2015ixa,Levi:2016ofk,Kim:2021rfj,Levi:2020uwu,Cho:2022syn,Kim:2022bwv}.

However, an excellent alternative approach to the bound two-body problem comes by way of studying two-body scattering:
here it is natural to define gauge-invariant scattering observables in terms
of the states at past-/future-infinity, where the gravitational field is weak.
It is also natural here to adopt the post-Minkowskian (PM) expansion in Newton's constant $G$,
which resums terms from infinitely high velocities in the post-Newtonian (PN) series.
One may use analytic continuation to directly produce PM observables for bound orbits
\cite{Kalin:2019rwq,Kalin:2019inp,Saketh:2021sri,Cho:2021arx};
alternatively, conservative scattering observables may be used to infer a Hamiltonian for the two-body system
\cite{Bjerrum-Bohr:2013bxa,Cheung:2018wkq,Neill:2013wsa,
  Vaidya:2014kza,Damour:2017zjx,Bjerrum-Bohr:2019kec,Cristofoli:2020uzm}.
A more sophisticated version of this strategy is to infer an effective-one-body (EOB) Hamiltonian
\cite{Buonanno:1998gg,Damour:2008yg,Damour:2019lcq,Antonelli:2019ytb,Damgaard:2021rnk},
which may be extended to include spin
\cite{Vines:2016unv,Vines:2017hyw,Vines:2018gqi,Bini:2017xzy,Bini:2018ywr}
and resums information from the test-body limit.

The Worldline Quantum Field Theory (WQFT) is a new formalism for producing gravitational scattering observables
\cite{Mogull:2020sak,Jakobsen:2021smu,Jakobsen:2021lvp,Jakobsen:2021zvh,Shi:2021qsb,Bastianelli:2021nbs,Jakobsen:2022fcj,Wang:2022ntx,Jakobsen:2022psy}.
It builds on the highly successful PM-based worldline EFT approach~\cite{Kalin:2020mvi},
which has been used to produce scattering observables at 3PM~\cite{Kalin:2020fhe,Kalin:2020lmz,Kalin:2022hph}
and 4PM orders~\cite{Dlapa:2021npj,Dlapa:2021vgp,Dlapa:2022lmu,Khalil:2022ylj};
the worldline EFT has also produced gravitational Bremsstrahlung
and radiative observables including tidal effects and spin
\cite{Mougiakakos:2021ckm,Riva:2021vnj,Mougiakakos:2022sic,Riva:2022fru}.
The WQFT goes a step further by quantizing worldline degrees of freedom,
which bypasses the need for intermediate off-shell objects such as the effective action.
A supersymmetric extension to the worldline accounts for quadratic-in-spin effects
\cite{Jakobsen:2021lvp,Jakobsen:2021zvh},
conveniently avoiding the typical use of a body-fixed frame.
In \Rcite{Jakobsen:2022fcj} we used the WQFT to produce conservative scattering observables ---
the momentum impulse $\Delta p_i^\mu$ and spin kick $\Delta S_i^\mu$ --- at 3PM order.

In this paper, we upgrade these observables to include radiation-reaction (dissipative) effects,
using the Schwinger-Keldysh in-in formalism~\cite{Schwinger:1960qe,Keldysh:1964ud,Jordan:1986ug,Weinberg:2005vy,Galley:2009px}
that has recently been incorporated into both the WQFT and PM-based worldline EFT frameworks~\cite{Jakobsen:2022psy,Kalin:2022hph}.
Our results confirm the radiated four-momentum $P_{\rm rad}^\mu$ recently
predicted with the worldline EFT approach~\cite{Riva:2022fru}.
Given these new observables, we postulate and confirm an extension to
Bini and Damour's linear response relation~\cite{Bini:2012ji,Damour:2020tta,Bini:2021gat}
which allows us to predict terms in the conservative and radiative parts of the full
scattering observables, depending on their behavior under the time-reversal operation $v_i^\mu\to-v_i^\mu$.
The extension holds for arbitrary spin orientations, and goes beyond linear response.

The WQFT is inspired by QFT amplitudes-based methods for tackling the classical two-body problem
\cite{Bjerrum-Bohr:2018xdl,Bjerrum-Bohr:2022blt,Kosower:2022yvp,Buonanno:2022pgc}.
These build on well-honed strategies for deriving scattering amplitudes \cite{Dixon:1996wi,Elvang:2013cua,Henn:2014yza,Bern:2019prr,Travaglini:2022uwo}
and performing the associated loop integrals \cite{Weinzierl:2022eaz,Blumlein:2022qci}.
In the non-spinning case a slew of two-body results have been produced at 
3PM order (two loops)~\cite{Bern:2019nnu,Bern:2019crd,Bern:2020gjj,Cheung:2020gyp,DiVecchia:2020ymx,Cheung:2020sdj,Bern:2020uwk,AccettulliHuber:2020dal}
and at 4PM order (three loops)~\cite{Bern:2021dqo,Bern:2021yeh,Khalil:2022ylj}.
There has also recently been work on $N$-body scattering and potentials~\cite{Jones:2022aji}.
Radiation-reaction effects have been incorporated
\cite{Herrmann:2021lqe,Herrmann:2021tct,DiVecchia:2021ndb,DiVecchia:2021bdo,Heissenberg:2021tzo,Bjerrum-Bohr:2021din,Damgaard:2021ipf},
and an in-in style formalism for directly producing observables has been introduced
\cite{Kosower:2018adc,Maybee:2019jus,Cristofoli:2021vyo,Cristofoli:2021jas}.
To handle spin, higher-spin fields are used
\cite{Guevara:2018wpp,Bautista:2019tdr,Guevara:2019fsj,Arkani-Hamed:2019ymq,Guevara:2020xjx,Bautista:2021wfy,Chiodaroli:2021eug,Aoude:2021oqj,Bautista:2021inx,Cangemi:2022abk}
and results have been produced at 2PM order at quadratic
\cite{Vines:2018gqi,Bern:2020buy,Kosmopoulos:2021zoq},
quartic~\cite{Chen:2021kxt} and higher orders in spin~\cite{Bern:2022kto,Aoude:2022trd,Aoude:2022thd}.
Similar results have also been achieved with the closely related heavy-particle EFT
\cite{Damgaard:2019lfh,Aoude:2020onz,Brandhuber:2021kpo,
  Brandhuber:2021eyq,Aoude:2020ygw,Haddad:2021znf}.

Most notably, a 3PM quadratic-in-spin Hamiltonian has now been derived using amplitudes-based methods~\cite{FebresCordero:2022jts},
involving spin on one of the two massive bodies only and without finite-size corrections.
In this paper, using the conservative scattering observables derived in \Rcite{Jakobsen:2022fcj}
we both confirm this result and extend it to include spin-spin effects and finite-size corrections relevant for neutron stars.
Quite remarkably, we find that knowledge of a single scattering angle suffices to completely determine the Hamiltonian,
also when arbitrarily mis-aligned spin vectors are involved.

Our paper is structured as follows.
In \Sec{sec:spinningCompactBodies} we review the dynamics of spinning massive bodies,
including their description up to quadratic order in spin in terms of an $\cN=2$ supersymmetric worldline action.
We demonstrate how, with a suitable SUSY shift, we can switch between the canonical and covariant
spin-supplementary conditions (SSCs).
In \Sec{sec:wqftInIn} we review the Schwinger-Keldysh in-in formalism in the context of WQFT,
and in \Sec{sec:radObservables} put it to use deriving the complete 3PM quadratic-in-spin momentum impulse $\Delta p_1^\mu$
and spin kick $\Delta S_1^\mu$ including radiation-reaction effects.
We present the results schematically, demonstrate how one may introduce scattering angles for mis-aligned spins,
and perform various consistency checks.

Next, in \Sec{sec:linResponse} we upgrade the linear response relation to mis-aligned spin directions,
generating both conservative and radiative terms from the full 3PM scattering observables $\Delta p_1^\mu$ and $\Delta S_1^\mu$.
In \Sec{sec:ham} we use the conservative scattering observables,
and in particular the scattering angle, to build a complete 3PM quadratic-in-spin Hamiltonian.
Finally, in \Sec{sec:binding} we discuss unbound-to-bound mappings for the specific case of aligned spins:
we generate the binding energy for circular orbits,
both numerically and analytically and up to 4PN order,
and produce plots of the binding energy as a function of the orbital frequency close to merger ---
comparing our results with numerical relativity.
We also determine the energy radiated per orbit using an appropriate analytic continuation~\cite{Cho:2021arx}.
In \Sec{sec:conclusions} we conclude.

\section{Spinning massive bodies}\label{sec:spinningCompactBodies}

A pair of black holes or neutron stars interacting through
a gravitational field in $D$-dimensional Einstein gravity are described by
\begin{align}\label{eq:actionComplete}
  S&=S_{\rm EH}[g_{\mu\nu}]+S_{\rm gf}[g_{\mu\nu}]
  +\sum_{i=1}^2S^{(i)}[g_{\mu\nu},x_i^\mu,\psi_i^a]\,,
\end{align}
where $S_{\rm EH}$ is the Einstein-Hilbert action ($\kappa=\sqrt{32\pi G}$),
\begin{align}\label{eq:SEH}
  S_{\rm EH}&=-\frac2{\kappa^2}\int\!\d^Dx\sqrt{-g}\,R\,,
\end{align}
$S_{\rm gf}$ is a gauge-fixing term
and $S^{(i)}$ are the two worldline actions.
Up to quadratic order in spin~\cite{Jakobsen:2021lvp,Jakobsen:2021zvh}
\begin{align}\label{eq:action}
    \frac{S^{(i)}}{m_i}&=-\!\int\!\d\tau_i \Big[\sfrac{1}{2}g_{\mu\nu}\dot x_i^{\mu}\dot x_i^{\nu}
      \!+\! i\bar\psi_{i,a}\!\sfrac{\D\psi_i^a}{\D\tau_i}\!+\!\sfrac12
      R_{abcd}\bar\psi_i^{a}\psi_i^{b}\bar\psi_i^{c}\psi_i^{d}\nn\\
      &\qquad\qquad +C_{{\rm E},i}E_{i,ab}
      \bar\psi_i^a \psi_i^b P_{i,cd}\bar\psi_i^{c}\psi_i^{d}\,\Big]\,,
  \end{align}
  where the projector is
  $P_{i,ab}:=\eta_{ab}-e_{a\mu}e_{b\nu}\dot{x}_i^\mu\dot{x}_i^\nu/\dot{x}_i^2$,
  $\eta_{ab}$ is the (mostly-minus) Minkowski metric
  and $E_{i,ab}:=R_{a\mu b\nu}\dot{x}_i^\mu\dot{x}_i^\nu/\dot{x}_i^2$.
The finite-size multipole moment coefficients $C_{{\rm E},i}$
are defined such that $C_{{\rm E},i}=0$ for black holes, and
\begin{align}
  \frac{\D\psi_i^a}{\D\tau_i}:=\dot{x}_i^\mu\nabla_\mu\psi_i^a=
  \dot\psi_i^a+\dot x_i^\mu{\omega_\mu}^{ab}\psi_{i,b}\,.
\end{align}
The two bodies with masses $m_i$ have positions $x_i^\mu(\tau_i)$;
the complex anticommuting fields $\psi_i^a(\tau_i)$,
defined in a local frame $e^a_\mu(x)$ with $g_{\mu\nu}=e_\mu^a e_\nu^b\eta_{ab}$,
encode spin degrees of freedom.

The worldline action~\eqref{eq:action} enjoys a global $\cN=2$ supersymmetry:
\be
\label{N=2SUSY}
\delta x_i^{\mu} = i\bar\eps_i \psi_i^{\mu} + i\eps_i \bar\psi_i^{\mu} \,,\,\,
\delta \psi_i^{a}= -\eps_i e^{a}_{\mu}{\dot x}_i^{\mu} -\delta x_i^{\mu}\, \omega_{\mu}{}^{a}{}_{b}\psi_i^{b},
\ee
with constant SUSY parameters $\eps_i$ and $\bar \eps_i = \eps_i^{\dagger}$.
As shown in \Rcite{Jakobsen:2021zvh},
these shifts are generated by the conserved supercharges
$\dot{x}_i\cdot\psi_i$ and $\dot{x}_i\cdot\bar\psi_i$.
There is also a U(1) symmetry:
\begin{align}
  \delta\psi_i^a&=i\alpha_i\psi_i^a\,,&
  \delta\bar\psi_i^a&=-i\alpha_i\bar\psi_i^a\,,&
  \delta x_i^\mu&=0\,,
\end{align}
generated by the conserved charge $\psi_i\cdot\bar\psi_i$.
Lastly, reparametrization invariance of the worldlines in $\tau_i$ implies
  \begin{equation}
    \dot{x}_i^2=1+R_{abcd}\bar\psi_i^a\psi_i^b\bar\psi_i^c\psi_i^d
    +2C_{{\rm E},i}E_{i,ab}\bar\psi_i^a \psi_i^b P_{i,cd}\bar\psi_i^{c}\psi_i^{d}
  \end{equation}
is also preserved.
As $\dot{x}_i^2\neq1$ generically along the worldlines this implies that $\tau_i$
are not the proper times; however, as we are generally only interested
in the asymptotic behavior this subtlety will not be important.

\subsection{Background Symmetries}

Fields are perturbatively expanded around their background values at past infinity:
\begin{subequations}\label{eq:perturbs}
  \begin{align}
    g_{\mu\nu}(x)&=\eta_{\mu\nu}+\kappa h_{\mu\nu}(x)\,,\\
    x_i^\mu(\tau_i)&=b_i^\mu+\tau_i v_i^\mu+z_i^\mu(\tau_i)\,,\\
    \psi_i^a(\tau_i)&=\Psi_i^a+\psi_i^{\prime a}(\tau_i)\,,
  \end{align}
\end{subequations}
where $p_i^\mu=m_iv_i^\mu$ is the initial momentum;
the initial value of the spin tensor is given by
\begin{equation}\label{eq:spinTensor}
  S_i^{ab}=-2im_i\bar\Psi_i^{[a}\Psi_i^{b]}\,.
\end{equation}
The antisymmetrization $[ab]$ includes a factor $1/2$ ---
note that this normalization of the spin tensor differs from that used in \Rcites{Jakobsen:2021lvp,Jakobsen:2021zvh,Jakobsen:2022fcj}.
The vierbein is similarly expanded as
\be
e^{a}_{\mu} = \eta^{a\nu}\left(\eta_{\mu\nu}+ \frac{\kappa}{2}h_{\mu\nu} - 
\frac{\kappa^2}{8}h_{\mu\rho}{h^\rho}_\nu  + \cO(\kappa^3) \right)\, ,
\ee
which allows us to drop the distinction between spacetime $\mu,\nu,\ldots$
and local frame $a,b,\ldots$ indices.
The global $\cN=2$ SUSY in the far past is
\begin{align}\label{susybg}
  \delta b^{\mu}_{i}&= i \bar\eps_i \Psi^{\mu}_{i} + i \eps_i\bar{\Psi}^{\mu}_{i}\, ,
  \quad \delta v^{\mu}_{i}=0 \, ,\nn
  \quad \delta\Psi_{i}^\mu=-\eps_i v_{i}^{\mu}\,, \\
  \Rightarrow\quad \delta S_{i}^{\mu\nu} &= 2p_{i}^{[\mu}\, \delta b_{i}^{\nu]} \,.
\end{align}
To fix these symmetries we find it convenient to
enforce the covariant spin-supplementary condition (SSC):
\begin{align}\label{eq:covSSC}
  p_i\cdot\Psi_i=0\quad\implies\quad p_{i,\mu}S_i^{\mu\nu}=0\,,
\end{align}
Using the reparametrization symmetry we also enforce
$v_i^2=1$ and $b\cdot v_i=0$, where $b^\mu=b_2^\mu-b_1^\mu$
is the impact parameter pointing from the first to the second massive body.
Finally, $\gamma=v_1\cdot v_2$;
we will also make use of unit-normalized ``hatted'' variables,
e.g.~$\hat{b}^\mu=b^\mu/|b|$.

The total initial angular momentum of the system is
\begin{align}
  \begin{aligned}
    J^{\mu\nu}&=L^{\mu\nu}+S_1^{\mu\nu}+S_2^{\mu\nu}\,,\\
    L^{\mu\nu}&=2b_1^{[\mu}p_1^{\nu]}+2b_2^{[\mu}p_2^{\nu]}\,,
  \end{aligned}
\end{align}
where $L^{\mu\nu}$ is the orbital component.
In this context, we see that the background symmetries
\eqref{susybg} correspond simply to invariance of the system's
total angular momentum under shifts in the origins of the two bodies $b_i^\mu$.
We can also shift the center of our coordinate system $x^\mu\to x^\mu+a^\mu$,
in which case $L^{\mu\nu}\to L^{\mu\nu}+2a^{[\mu}P^{\nu]}$ as discussed in \eg~\Rcite{Manohar:2022dea}.
The orbital and spin angular momentum vectors,
defined specifically in $D=4$ dimensions,
are invariant under these shifts:
\begin{subequations}
  \begin{align}
    L^\mu&:=\sfrac12{\eps^\mu}_{\nu\rho\sigma}L^{\nu\rho}\hat P^\sigma=
    -\sfrac{1}{E} {\eps^\mu}_{\nu\rho\sigma} b^\nu p_1^\rho p_2^\sigma\,,\label{eq:Lvec}\\
    S_i^\mu&:=m_ia_i^\mu=\sfrac12{\eps^\mu}_{\nu\rho\sigma}S_i^{\nu\rho}v_i^\sigma\,,\label{eq:aVec}
  \end{align}
\end{subequations}
where
\begin{subequations}
  \begin{align}
    P^\mu&=p_1^\mu+p_2^\mu\,,\\
    p^\mu&=\frac{m_1m_2}{E^2}\big[(\gamma m_1+m_2)v_1^\mu-(\gamma m_2+m_1)v_2^\mu\big]\,,\label{eq:comMom}
  \end{align}
\end{subequations}
are respectively the total and center-of-mass (CoM) momentum, $p^\mu=(0,\mathbf{p}_\infty)$.
Here $E=|P|=M\,\Gamma=M\sqrt{1+2\nu(\gamma-1)}$ is the energy in the CoM frame,
$M=m_1+m_2$, $\nu=\mu/M=m_1m_2/M^2$ are the total mass and symmetric mass ratio;
$p_\infty=|\mathbf{p}_\infty|=\mu\sqrt{\gamma^2-1}/\Gamma$ is the center-of-mass momentum.
With the covariant SSC choice~\eqref{eq:covSSC}
the total angular momentum $J^\mu$ is given by
\begin{align}\label{eq:jVec}
  \begin{aligned}
    J^\mu:=&\sfrac12{\eps^\mu}_{\nu\rho\sigma}J^{\nu\rho} \hat P^\sigma\\
    =&L^\mu+\sum_i\left(v_i\cdot \hat P\,S_i^\mu-S_i\cdot \hat P\,v_i^\mu\right)\,.
  \end{aligned}
\end{align}
Notice that $J^\mu\neq L^\mu+S_1^\mu+S_2^\mu$,
which is due to $S_i^\mu$ being defined in their
respective inertial frames $v_i^\mu$
rather than the center-of-mass frame $\hat{P}^\mu$.

\subsection{Canonical Spin Variables}
\label{sec:canonical}

We also find it useful to introduce
canonical variables \cite{Barausse:2009aa,Vines:2016unv,Vines:2017hyw}
which are designed to ensure that
\begin{align}
  J^\mu=L_{\rm can}^\mu+S_{1,\rm can}^\mu+S_{2,\rm can}^\mu\,,
\end{align}
and $P\cdot L_{\rm can}=P\cdot S_{i,\rm can}=0$.
The canonical spin vectors $S_{i,\rm can}^\mu$ are given by a boost
of the covariant spin vectors $S_i^\mu$ to the center-of-mass frame:
\begin{align}
  \begin{aligned}\label{eq:canonical}
    S_{i,\rm can}^\mu&:={\Lambda^\mu}_\nu(v_i\to\hat{P})S^\nu_i\\
    &=S_i^\mu-\frac{\hat{P}\cdot S_i}{\gamma_i+1}(\hat{P}^\mu+v_i^\mu)\,,
  \end{aligned}
\end{align}
where $\gamma_i=\hat{P}\cdot v_i$ is the time component of $v_i^\mu$
in the center-of-mass frame.
To ensure preservation of the total angular momentum $J^\mu$~\eqref{eq:jVec},
we have
\begin{align}\label{eq:lShift}
  L_{\rm can}^\mu&=L^\mu+\sum_{i=1}^2\bigg[(\gamma_i\!-\!1)S_i^\mu\!+\!\frac{\hat{P}\cdot S_i}{\gamma_i\!+\!1}(\hat{P}^\mu-\gamma_iv_i^\mu)\bigg].
\end{align}
The canonical impact parameter $b_{\rm can}^\mu$ ---
in terms of which
${L_{\rm can}^\mu=-E^{-1}{\eps^\mu}_{\nu\rho\sigma}b_{\rm can}^\nu p_1^\rho p_2^\sigma}$ ---
is related to $b^\mu$ by a specific SUSY shift~\eqref{susybg}:
\begin{align}
  \eps_i=\frac{\hat{P}\cdot\Psi_i}{\gamma_i+1}\,.
\end{align}
We then have, with $E_i=\gamma_i m_i$
\begin{subequations}
  \begin{align}
    b^\mu_{i,\rm can}&=b_i^\mu+\frac1{E_i+m_i}S_i^{\mu\nu}\hat{P}_\nu\,,\\
    \Psi^\mu_{i,\rm can}&=\Psi^\mu_i-\frac{\hat{P}\cdot\Psi_i}{\gamma_i+1}v_i^\mu\,,
  \end{align}
\end{subequations}
and can confirm that the the canonical spin tensor
$S^{\mu\nu}_{i,\rm can}=-2im_i\bar\Psi_{i,\rm can}^{[\mu}\Psi_{i,\rm can}^{\nu]}$
satisfies the canonical
Pryce-Newton-Wigner SSC~\cite{Pryce:1935ibt,Pryce:1948pf,Newton:1949cq}:
\begin{align}
  (\hat{P}+v_i)\cdot\Psi_{i,\rm can}=0 \implies
  (\hat{P}_\mu+v_{i,\mu})S^{\mu\nu}_{i,\rm can}=0\,.
\end{align}
The canonical spin vector is then also given by
\begin{align}\label{eq:canonical2}
  S^\mu_{i,\rm can}&=
  \sfrac12{\eps^\mu}_{\nu\rho\sigma}S^{\nu\rho}_{i,\rm can}\hat{P}^\sigma\,,
\end{align}
and has a vanishing time component in the center-of-mass frame:
$P\cdot S_{i,\rm can}=0$.
This will be useful when we construct a Hamiltonian in \Sec{sec:ham}.

\section{WQFT In-In formalism}\label{sec:wqftInIn}

Complete observables including both conservative and radiative contributions
are produced from WQFT using the Schwinger-Keldysh in-in formalism
\cite{Schwinger:1960qe,Keldysh:1964ud,Jordan:1986ug,Weinberg:2005vy,Galley:2009px}.
Formally this involves doubling the degrees of freedom in our theory:
$h_{\mu\nu}\to\{h_{1\mu\nu},h_{2\mu\nu}\}$,
$z_i^\mu\to\{z_{1i}^\mu,z_{2i}^\mu\}$ and
$\psi_i^{\prime\mu}\to\{\psi_{1i}^{\prime\mu},\psi_{2i}^{\prime\mu}\}$.
Observables are defined in terms of a path integral including two copies of the action:
\begin{align}\label{eq:expValue}
  &\braket{\cO}_{\rm in-in}
  :=\int\!{\cal D}[h_{A\mu\nu},z_{Ai}^\mu,{\psi_{Ai}^{\prime\mu}}]
  e^{i(S[\{\}_1]-S^*[\{\}_2])}\cO\,,
\end{align}
where $A=1,2$ and we use the shorthand
$\{\}_A:=\{g_{A\mu\nu},x_{Ai}^\mu,{\psi_{Ai}^{\mu}}\}$.
The boundary conditions on $h_{A\mu\nu}$, $z_{Ai}^\mu$ and $\psi_{Ai}^{\prime\mu}$
are that all fields equate at future infinity,
\begin{subequations}
  \begin{align}
    h_{1\mu\nu}(t=+\infty,\mathbf{x})&=h_{2\mu\nu}(t=+\infty,\mathbf{x})\,,\\
    z_{1i}^{\mu}(\tau_i=+\infty)&=z_{2i}^{\mu}(\tau_i=+\infty)\,,\\
    \psi_{1i}^{\prime\mu}(\tau_i=+\infty)&=\psi_{2i}^{\prime\mu}(\tau_i=+\infty)\,,
  \end{align}
\end{subequations}
and vanish at past infinity:
\begin{subequations}
  \begin{align}
    h_{1\mu\nu}(t=-\infty,\mathbf{x})&=h_{2\mu\nu}(t=-\infty,\mathbf{x})=0\,,\\
    z_{1i}^{\mu}(\tau_i=-\infty)&=z_{2i}^{\mu}(\tau_i=-\infty)=0\,,\\
    \psi_{1i}^{\prime\mu}(\tau_i=-\infty)&=\psi_{2i}^{\prime\mu}(\tau_i=-\infty)=0\,.
  \end{align}
\end{subequations}
This entangling of the boundary conditions gives rise to off-diagonal terms
in the propagator matrices involving the doubled fields.
For full details, see \Rcite{Jakobsen:2022psy}.

Fortunately, when performing calculations there
is no need to double degrees of freedom in this way.
The key insight of \Rcite{Jakobsen:2022psy} was that
tree-level single-operator expectation values~\eqref{eq:expValue} are produced using
precisely the same Feynman rules as in the in-out formalism,
but with retarded propagators pointing towards the outgoing line.
The retarded graviton propagator is
\begin{align}\label{eq:gravProp}
  \includegraphics[trim=0 .4cm 0 0]{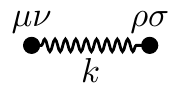}
  &=i\frac{P_{\mu\nu;\rho\sigma}}{k^2+{\rm sgn}(k^0)i0}\,,
\end{align}
where $P_{\mu\nu;\rho\sigma}:=\eta_{\mu(\rho}\eta_{\sigma)\nu}-
\sfrac1{D-2}\eta_{\mu\nu}\eta_{\rho\sigma}$
and $i0$ denotes a small positive imaginary part.
For the worldline modes $z_i^\mu$ and $\psi_i^{\prime\mu}$
the retarded propagators are respectively
\begin{subequations}\label{eq:wlPropagators}
  \begin{align}
    \includegraphics[trim=0 .3cm 0 0]{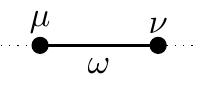}
    &\,=\,-i\frac{\eta^{\mu\nu}}{m_i(\omega+i0)^2}\,,\label{eq:propZ}
    \\
    \includegraphics[trim=0 .3cm 0 0]{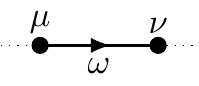}
    &\,=\,-i\frac{\eta^{\mu\nu}}{m_i(\omega+i0)}\,.\label{eq:propPsi}
  \end{align}
\end{subequations}
The Feynman vertices are unchanged with respect to the in-in formalism:
for example, the single-graviton emission vertex from worldline $i$ is
\begin{align}\label{eq:vertexH}
  \includegraphics[trim=0 1cm 0 0]{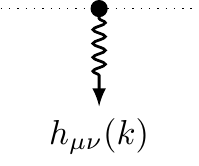}
  &=
  -i\frac{m_i\kappa }{2}e^{ik\cdot b_i}\dd(k\cdot v_i)
  \bigg(v_i^\mu v_i^\nu+ik_\rho S_i^{\rho(\mu}v_i^{\nu)} \nn \\
  &\!\!\!\!\!+\frac12k_\rho k_\sigma S_i^{\rho\mu} S_i^{\nu\sigma}
  +\frac{C_{{\rm E},i}}{2}v_i^\mu v_i^\nu k_\rho{{S_i}^\rho}_\sigma{{S_i}^\sigma}_\lambda k^\lambda\bigg),
\end{align}
where $\dd(\omega):=2\pi\delta(\omega)$.
At tree level the WQFT simply provides a diagrammatic mechanism
for solving the classical equations of motion in momentum space,
and so the use of retarded propagators ensures that boundary
conditions are fixed in the far past.

\section{Radiative observables}\label{sec:radObservables}

\begin{figure*}[t!]
  \centering
  \begin{subfigure}{0.12\textwidth}
    \centering
    \includegraphics{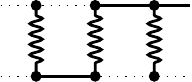}
    \caption{}
  \end{subfigure}
  \begin{subfigure}{0.12\textwidth}
    \centering
    \includegraphics{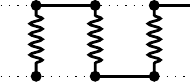}
    \caption{}
  \end{subfigure}
  \begin{subfigure}{0.12\textwidth}
    \centering
    \includegraphics{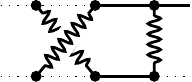}
    \caption{}
  \end{subfigure}
  \begin{subfigure}{0.12\textwidth}
    \centering
    \includegraphics{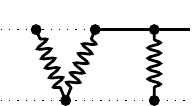}
    \caption{}
  \end{subfigure}
  \begin{subfigure}{0.12\textwidth}
    \centering
    \includegraphics{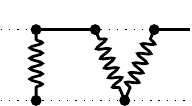}
    \caption{}
  \end{subfigure}
  \begin{subfigure}{0.12\textwidth}
    \centering
    \includegraphics{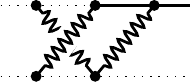}
    \caption{}
  \end{subfigure}
  \begin{subfigure}{0.12\textwidth}
    \centering
    \includegraphics{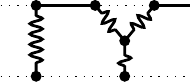}
    \caption{}
  \end{subfigure}
  \begin{subfigure}{0.12\textwidth}
    \centering
    \includegraphics{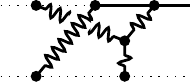}
    \caption{}
  \end{subfigure}
  \begin{subfigure}{0.12\textwidth}
    \centering
    \includegraphics{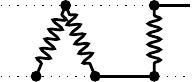}
    \caption{}
  \end{subfigure}
  \begin{subfigure}{0.12\textwidth}
    \centering
    \includegraphics{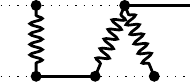}
    \caption{}
  \end{subfigure}
  \begin{subfigure}{0.12\textwidth}
    \centering
    \includegraphics{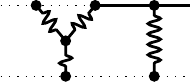}
    \caption{}
  \end{subfigure}
  \begin{subfigure}{0.12\textwidth}
    \centering
    \includegraphics{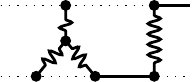}
    \caption{}
  \end{subfigure}
  \begin{subfigure}{0.12\textwidth}
    \centering
    \includegraphics{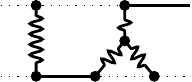}
    \caption{}
  \end{subfigure}
  \begin{subfigure}{0.12\textwidth}
    \centering
    \includegraphics{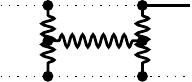}
    \caption{}
  \end{subfigure}
  \begin{subfigure}{0.12\textwidth}
    \centering
    \includegraphics{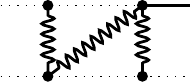}
    \caption{}
  \end{subfigure}
  \begin{subfigure}{0.12\textwidth}
    \centering
    \includegraphics{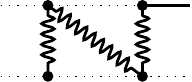}
    \caption{}
  \end{subfigure}
  \begin{subfigure}{0.12\textwidth}
    \centering
    \includegraphics{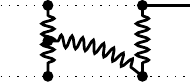}
    \caption{}
  \end{subfigure}
  \begin{subfigure}{0.12\textwidth}
    \centering
    \includegraphics{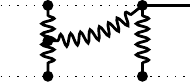}
    \caption{}
  \end{subfigure}
  \begin{subfigure}{0.12\textwidth}
    \centering
    \includegraphics{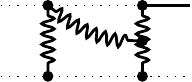}
    \caption{}
  \end{subfigure}
  \begin{subfigure}{0.12\textwidth}
    \centering
    \includegraphics{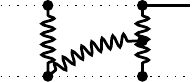}
    \caption{}
  \end{subfigure}
  \begin{subfigure}{0.12\textwidth}
    \centering
    \includegraphics{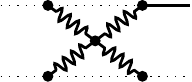}
    \caption{}
  \end{subfigure}
  \begin{subfigure}{0.12\textwidth}
    \centering
    \includegraphics{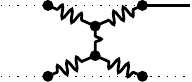}
    \caption{}
  \end{subfigure}
  \begin{subfigure}{0.12\textwidth}
    \centering
    \includegraphics{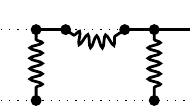}
    \caption{}
  \end{subfigure}
  \begin{subfigure}{0.12\textwidth}
    \centering
    \includegraphics{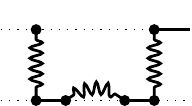}
    \caption{}
  \end{subfigure}
  \begin{subfigure}{0.12\textwidth}
    \centering
    \includegraphics{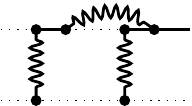}
    \caption{}
  \end{subfigure}
  \begin{subfigure}{0.12\textwidth}
    \centering
    \includegraphics{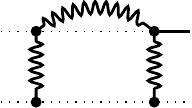}
    \caption{}
  \end{subfigure}
  \begin{subfigure}{0.12\textwidth}
    \centering
    \includegraphics{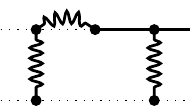}
    \caption{}
  \end{subfigure}
  \begin{subfigure}{0.12\textwidth}
    \centering
    \includegraphics{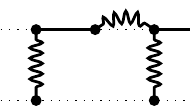}
    \caption{}
  \end{subfigure}
  \begin{subfigure}{0.12\textwidth}
    \centering
    \includegraphics{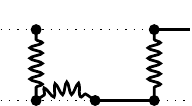}
    \caption{}
  \end{subfigure}
  \begin{subfigure}{0.12\textwidth}
    \centering
    \includegraphics{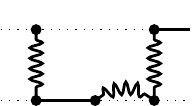}
    \caption{}
  \end{subfigure}
  \begin{subfigure}{0.12\textwidth}
    \centering
    \includegraphics{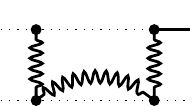}
    \caption{}
  \end{subfigure}
  \begin{subfigure}{0.12\textwidth}
    \centering
    \includegraphics{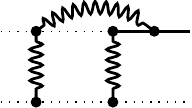}
    \caption{}
  \end{subfigure}
  \caption{\small
    The 32 types of diagrams contributing to the $m_1^2m_2^2$ components of
    $\Delta p_1^{(3)\mu}$ and the $m_1m_2^2$ components of $\Delta\psi_1^{(3)\mu}$.
    Diagrams (a)--(v) were already present in the conservative calculation~\cite{Jakobsen:2022fcj},
    though their expressions are modified by the inclusion of radiation;
    the mushrooms (w)--(ff) are purely radiative,
    and did not appear in the strictly conservative calculation~\cite{Jakobsen:2022fcj}.
    Diagram (y) includes the same worldline propagator with opposite $i0$ prescriptions,
    and so belongs to the $K$ integral family \eqref{eq:kIntegral}.
    For brevity we use solid lines to represent both propagating deflection $z_i^\mu$ and
    spin modes $\psi_i^{\prime\mu}$.
  }
  \label{fig:splitDiags}
\end{figure*}

Building on \Rcite{Jakobsen:2022fcj},
we compute the momentum impulse and change in $\psi_i^\mu$:
\begin{subequations}
  \begin{align}
    \Delta p_i^\mu&:=[m_i\dot{x}_i^\mu]^{\tau_i=+\infty}_{\tau_i=-\infty}=
    -m_i\omega^2\!\left.\braket{z_i^\mu(\omega)}_{\rm in-in}\right|_{\omega=0}\!,\\
    \Delta\psi_i^\mu&:=[\psi_i^\mu]^{\tau_i=+\infty}_{\tau_i=-\infty}=
    -i\omega\!\left.\braket{\psi_i^{\prime\mu}(\omega)}_{\rm in-in}\right|_{\omega=0}\,,
  \end{align}
\end{subequations}
but now also including radiation-reaction effects.
Using the definitions of the spin tensor $S_i^{\mu\nu}$~\eqref{eq:spinTensor}
and spin vector $S_i^\mu$~\eqref{eq:aVec}
we can then also derive the spin kick $\Delta S_i^\mu$:
\begin{align}\label{eq:spinKick}
  \Delta S_i^{\mu\nu}&=
  -2im_i\big(\bar\Psi_i^{[\mu}\Delta\psi_i^{\nu]}+
  \Delta\bar\psi_i^{[\mu}\Psi_i^{\nu]}+
  \Delta\bar\psi_i^{[\mu}\Delta\psi_i^{\nu]}\big)\,,\nn\\
  \Delta S_i^\mu\!&=\!\sfrac1{2m_i}{\eps^\mu}_{\nu\rho\sigma}\!\left(
  S_i^{\nu\rho}\Delta p_i^\sigma\!+\!\Delta S_i^{\nu\rho}p_i^\sigma\!+\!
  \Delta S_i^{\nu\rho}\Delta p_i^\sigma\right)\!.
\end{align}
We seek the 3PM components in a PM expansion:
\begin{align}\label{eq:pmDecomp}
  \Delta X=\sum_nG^n\Delta X^{(n)}\,,
\end{align}
where $\Delta X$ could be any of these observables:
$\Delta p_i^\mu$, $\Delta S_i^\mu$, $\Delta S_i^{\mu\nu}$ or $\Delta\psi_i^\mu$.

The relevant Feynman diagrams for both calculations are drawn in \Fig{fig:splitDiags}.
These diagrams make no distinction between conservative and radiative effects.
As only the $m_1^2m_2^2$ component of $\Delta p_1^{(3)\mu}$
and the $m_1m_2^2$ component of $\Delta\psi_1^{(3)\mu}$ are specifically affected
by the inclusion of radiation-reaction effects
we recompute these components; for the rest, we simply bring forwards
our previous results from \Rcite{Jakobsen:2022fcj}.
Integrands are assembled using the WQFT Feynman rules~\cite{Jakobsen:2021zvh}
in $D=4-2\eps$ dimensions,
which involves integration on the momenta or energies of all internal lines;
vertices contain either energy- or momentum-conserving $\dd$-functions, whichever is appropriate.
The energy integrals, corresponding to internal propagation of $z_i^\mu$ or $\psi_i^\prime$ modes, are trivial:
conservation of energy at the worldline vertices resolves them immediately.

Each graph has three unresolved four-momenta to integrate over.
The first of these integrals is a Fourier transform:
\begin{align}\label{eq:fourier}
  \begin{aligned}
    &\Delta X(b^\mu,v_i^\mu,S_i^{\mu\nu})\\
    &\qquad=\int_q e^{iq\cdot b}\dd(q\cdot v_1)\dd(q\cdot v_2)\Delta X(q^\mu,v_i^\mu,S_i^{\mu\nu})\,,
  \end{aligned}
\end{align}
where $q^\mu$ is the total momentum exchanged from the second to the first worldline,
and $\int_q:=\int\d^Dq/(2\pi)^D$.
Here we have implicitly defined the momentum-space observables $\Delta X(q^\mu,v_i^\mu,S_i^{\mu\nu})$,
which are given as linear combinations of two-loop Feynman integrals:
\begin{align}\label{eq:iIntegral}
  &\!\!\!\!\!\!\!\!\!
  I^{(\sigma_1;\sigma_2;\sigma_3)}_{n_1,n_2,\ldots,n_7}[\ell_1^{\mu_1}\cdots\ell_1^{\mu_n}\ell_2^{\nu_1}\cdots\ell_2^{\nu_n}]\nn\\
  :\!&=\int_{\ell_1,\ell_2}\!\frac{\dd(\ell_1\cdot v_2)\dd(\ell_2\cdot v_1)\ell_1^{\mu_1}\cdots\ell_1^{\mu_n}\ell_2^{\nu_1}\cdots\ell_2^{\nu_n}}
  {D_1^{n_1}D_2^{n_2}\cdots D_7^{n_7}}\,,\nn\\
  D_1&=\ell_1\cdot v_1+\sigma_1i0\,, \,
  D_2=\ell_2\cdot v_2+\sigma_2i0\,,\\
  D_3&=(\ell_1+\ell_2-q)^2+
  \sigma_3{\rm sgn}(\ell_1^0+\ell_2^0-q^0)i0\,,\nn\\
  D_4&=\ell_1^2\,, \,
  D_5=\ell_2^2\,,
  D_6=(\ell_1-q)^2\,,\,
  D_7=(\ell_2-q)^2\,.\nn
\end{align}
These integrals with retarded propagators were discussed at length in \Rcite{Jakobsen:2022psy}:
propagators $D_4$--$D_7$ are prevented from going on-shell by the
requirement that $\ell_1\cdot v_2=\ell_2\cdot v_1=0$,
so we can safely ignore their $i0$ prescriptions.
We also require
\begin{align}\label{eq:kIntegral}
  &\!\!\!\!\!\!\!\!\!
  K^{(\sigma)}_{n_1,n_2,n_3,n_4,n_5}[\ell^{\mu_1}\cdots\ell^{\mu_n}k^{\nu_1}\cdots k^{\nu_n}]\nn\\
  :\!&=\int_{\ell,k}\!\frac{\dd((k-\ell)\cdot v_1)\dd(\ell\cdot v_2)\ell^{\mu_1}\cdots\ell^{\mu_n}k^{\nu_1}\cdots k^{\nu_n}}
  {D_1^{n_1}D_2^{n_2}D_3^{n_3}D_4^{n_4}D_5^{n_5}}\,,\nn\\
  D_1&=\ell\cdot v_1+i0\,, \,
  D_2=\ell\cdot v_1-i0\,,\\
  D_3&=k^2+\sigma{\rm sgn}(k^0)i0\,,\,
  D_4=\ell^2\,, \,
  D_5=(\ell-q)^2\,,\nn
\end{align}
which accounts for the possibility of a worldline propagator appearing twice,
but with different $i0$ prescriptions ---
diagram (y) in \Fig{fig:splitDiags}.

The subsequent integration steps were discussed in \Rcite{Jakobsen:2022fcj},
and are not substantially different with the inclusion of radiation-reaction effects in the observables.
Tensorial two-loop integrals are reduced to scalar-type by expanding on a suitable basis,
and then reduced to master integrals using integration-by-parts (IBP) identities.
Expressions for these master integrals were provided in \Rcite{Jakobsen:2022psy},
and once the Fourier transform~\eqref{eq:fourier} has been performed on the exchanged momentum $q^\mu$
we are left with the observables in $D$ dimensions.
The scalar integrals themselves have simple reality properties:
\begin{align}\label{eq:integralsReality}
  \begin{aligned}
    I^{(\sigma_1;\sigma_2;\sigma_3)*}_{n_1,n_2,\ldots,n_7}
    &=(-1)^{n_1+n_2}I^{(\sigma_1;\sigma_2;\sigma_3)}_{n_1,n_2,\ldots,n_7}\,,\\
    K^{(\sigma)*}_{n_1,n_2,\ldots,n_5}
    &=(-1)^{n_1+n_2}K^{(\sigma)}_{n_1,n_2,\ldots,n_5}\,,
  \end{aligned}
\end{align}
i.e.~they are either purely real or imaginary,
depending on whether they have an even or odd number of worldline propagators respectively.
While this implies that the momentum-space observables $\Delta X(q^\mu,v_i^\mu,S_i^{\mu\nu})$
are complex functions, the Fourier transform~\eqref{eq:fourier} introduces additional factors of $i$,
giving rise to purely real observables $\Delta X(b^\mu,v_i^\mu,S_i^{\mu\nu})$.

The final observables $\Delta p_i^{(3)\mu}$ and $\Delta\psi_i^{(3)\mu}$
in four dimensions are found by taking the limit $D\to4$,
checking to ensure the cancellation of all poles in
the dimensional regularization parameter $\eps=2-\frac{D}2$.
We generated our integrand in $D=4-2\eps$ dimensions
in order to account for possible cancellations of $\eps$ poles with the two-loop integrals.
Like in \Rcite{Jakobsen:2022fcj} we have verified that the supercharges
$p_i^2$, $\psi_i\cdot\bar\psi_i$, $p_i\cdot\psi_i$ and $p_i\cdot\bar\psi_i$ are conserved.
This means that the following identities:
\begin{align}\label{eq:checks}
  &0=p_1\!\cdot\!\Delta p_1^{(3)}\!+\!\Delta p_{1}^{(1)}\!\cdot\!\Delta p_1^{(2)}\,,\\
  &0=\bar\Psi_1\!\cdot\!\Delta\psi_1^{(3)}\!+\!\Delta\bar\psi_1^{(3)}\!\cdot\!\Psi_1\!+\!
  \Delta\bar\psi_1^{(1)}\!\cdot\!\Delta\psi_1^{(2)}\!+\!
  \Delta\bar\psi_1^{(2)}\!\cdot\!\Delta\psi_1^{(1)},\nn\\
  &0=p_1\!\cdot\!\Delta\psi_1^{(3)}\!+\!\Delta p^{(3)}_1\!\cdot\!\Psi_1\!+\!
  \Delta p_1^{(1)}\!\cdot\!\Delta\psi_1^{(2)}\!+\!
  \Delta p_1^{(2)}\!\cdot\!\Delta\psi_1^{(1)}\!\!.\nn
\end{align}
are satisfied.

\subsection{Results}

We find it convenient to decompose the observables into four gauge-invariant parts each:
\begin{subequations}
  \begin{align}
    \Delta p_i^\mu&=\Delta p_{i,\rm cons}^{(+)\mu}+\Delta p_{i,\rm cons}^{(-)\mu}+
    \Delta p_{i,\rm rad}^{(+)\mu}+\Delta p_{i,\rm rad}^{(-)\mu}\,,\label{eq:parityP}\\
    \Delta S_i^\mu&=\Delta S_{i,\rm cons}^{(+)\mu}+\Delta S_{i,\rm cons}^{(-)\mu}+
    \Delta S_{i,\rm rad}^{(+)\mu}+\Delta S_{i,\rm rad}^{(-)\mu}\,.\label{eq:parityA}
  \end{align}
\end{subequations}
The split into conservative `cons' and radiative `rad' pieces is done with respect to the integrals
\eqref{eq:iIntegral}, \eqref{eq:kIntegral}:
the potential and radiative regions~\cite{Jakobsen:2022psy}.
Meanwhile the split into $(\pm)$ sectors is defined 
with respect to behavior under a time-reversal operation:
\begin{align}\label{eq:timeReversal}
  \left.\Delta X^{(\pm)}\right|_{v_i^\mu\to-v_i^\mu}&=\pm\Delta X^{(\pm)}\,,
\end{align}
which flips the signs on the timelike vectors $v_i^\mu$
(and the momenta $p_i^\mu=m_i v_i^\mu$) but
not the spacelike vectors $b^\mu$ and $a_i^\mu$,
leaving $\gamma=v_1\cdot v_2$ invariant.
Under this operation,
$S_i^{\mu\nu}={\eps^{\mu\nu}}_{\rho\sigma}p_i^\rho a_i^\sigma$ changes sign.

For the impulse, we have
\begin{align}
  \Delta p_{1,\rm cons}^{(3;+)\mu}
  &=
  \frac{m_1^2 m_2^2}{|b|^3}
  \Big[
    c_1^{(+)\mu}
    \frac{\text{arccosh}\gamma}{\sqrt{\gamma^2-1}}
    \!+\!
    \sum_{n=1}^3
    \Big(
    \frac{m_1}{m_2}
    \Big)^{n-2}\!
    c_{n+1}^{(+)\mu}
    \Big],\nn\\
  \Delta p_{1,\rm rad}^{(3;+)\mu}
  &=
  \frac{m_1^2 m_2^2}{|b|^{3}}{\mathcal I}(v)
  c_5^{(+)\mu}\nn\\
  \Delta p_{1,\rm cons}^{(3;-)\mu}
  &=
  \sum_{n=1}^3
  \frac{\pi m_1^2 m_2^2}{|b|^{3}}
  \Big(
  \frac{m_1}{m_2}\!
  \Big)^{n-2}
  c_{n}^{(-)\mu}\,,\\
  \Delta p_{1,\rm rad}^{(3;-)\mu}
  &=
  \frac{\pi m_1^2 m_2^2}{|b|^{3}}\times\nn\\
  &\quad\Big[
    c_{4}^{(-)\mu} +
    c_{5}^{(-)\mu}
    \frac{\text{arccosh}\gamma}{\sqrt{\gamma^2-1}}
    \!+\!
    c_{6}^{(-)\mu}
    \log\left(\frac{1+\gamma}2\right)\Big],\nn
\end{align}
where
\begin{align}\label{eq:univPrefactor}
  {\cal I}(v)=-\frac83+\frac1{v^2}+\frac{(3v^2-1)}{v^3}{\rm arccosh}(\gamma)
\end{align}
is a universal prefactor, and $v=\sqrt{\gamma^2-1}/\gamma$.
The vectors are given by
\begin{align}
  c_n^{(\pm)\mu}&=\vec{f}_n^{\,\,(\pm)}(\gamma,C_{{\rm E},i})\cdot{\vec{\rho}_\pm}^{\,\mu}\,,
\end{align}
with basis elements even/odd under time reversal:
\begin{subequations}\label{eq:basisVectors}
  \begin{align}
    {\vec{\rho}_+}^{\,\mu}&=\bigg\{
    \hat{b}^\mu,
    \frac{a_i\cdot\hat{L}}{|b|}\hat{b}^\mu,
    \frac{a_i\cdot\hat{b}}{|b|}\hat{L}^\mu,
    \frac{a_i\cdot a_j}{|b|^2}\hat{b}^\mu,
    \frac{a_i\cdot\hat{b}\,a_j\cdot\hat{b}}{|b|^2}\hat{b}^\mu,\nn\\
    &\!\!\!\!
    \frac{a_i\cdot v_{\bar\imath}\,a_j\cdot v_{\bar\jmath}}{|b|^2}\hat{b}^\mu,
    \frac{a_i \cdot\hat{b}\,a_j\cdot \hat{L}}{|b|^2}\hat{L}^\mu,
    \frac{a_i\cdot \hat{b}\,a_j\cdot v_{\bar\jmath}}{|b|^2}v_k^\mu
    \bigg\}\,,\\
	     {\vec{\rho}_-}^{\,\mu}&=\bigg\{
	     v_i^\mu,
	     \frac{a_i\cdot \hat{L}}{|b|}v_j^\mu,
	     \frac{a_i\cdot v_{\bar\imath}}{|b|}\hat{L}^\mu,
	     \frac{a_i\cdot a_j}{|b|^2}v_k^\mu,
	     \frac{a_i\cdot v_{\bar\imath}\,a_j\cdot v_{\bar\jmath}}{|b|^2}v_k^\mu
             ,\nn\\
	     &\!\!\!\!
	     \frac{a_i\cdot\hat{b}\,a_j\cdot\hat{b}}{|b|^2}v_k^\mu,
	     \frac{a_i\cdot v_{\bar\imath}\,a_j\cdot\hat{L}}{|b|^2}\hat{L}^\mu,
	     \frac{a_i\cdot v_{\bar\imath}\,a_j\cdot \hat{b}}{|b|^2}\hat{b}^\mu
	     \bigg\},
  \end{align}
\end{subequations}
where $i,j,k=1,2$ and $\bar{1}=2$, $\bar{2}=1$,
$\hat{L}^\mu=L^\mu/|L|$ and $\hat{b}^\mu=b^\mu/|b|$ being unit-normalized vectors.
Except for denominators of $(\gamma^2-1)^{n/2}$ the
remaining scalar components $\vec{f}_n^{\,\,(\pm)}$
are polynomials in $\gamma$, $C_{{\rm E},i}$,
so we refrain from providing explicit expressions in the text;
instead, we
refer the reader to the ancillary file attached to the \texttt{arXiv}
submission of this paper for full expressions.

Let us remark on certain properties of this result.
$\Delta p_i^{(3;+)\mu}$ and $\Delta p_i^{(3;-)\mu}$ are respectively associated with the
real (imaginary) integrals~\eqref{eq:integralsReality},
i.e.~those with an even (odd) number of worldline propagators.
The factors of $\pi$ in $\Delta p_i^{(3;-)\mu}$ thus arise from the overall factors of $i\pi$
in the purely imaginary master integrals --- see \Rcite{Jakobsen:2022psy}.
We also note the behavior under $v\to -v$:
in each case, the radiative components pick up the opposite sign from the conservative components.
Finally, the function ${\cal I}(v)$ is familiar:
it appears in the 2PM radiated angular momentum~\eqref{eq:jRad}.
As we shall see in \Sec{sec:linResponse},
$\Delta p_{i,\rm rad}^{(3;+)\mu}$ and $\Delta p_{i,\rm cons}^{(3;-)\mu}$ can
be inferred directly from lower-PM observables,
using a generalization to the linear response relation
\cite{Bini:2012ji,Damour:2020tta,Bini:2021gat}.

From the impulse $\Delta p_i^\mu$ we straightforwardly recover
the four-momentum radiated from the scattering event:
\begin{equation}\label{eq:radP}
  P^\mu_{\rm rad}=-\Delta P^\mu=-\Delta p_1^\mu-\Delta p_2^\mu\,,
\end{equation}
which vanishes if we consider only conservative scattering.
Here we agree with a recent 3PM worldline EFT result for $P^\mu_{\rm rad}$
obtained by Riva, Vernizzi and Wong~\cite{Riva:2022fru}.
We also agree with our own previous result for the leading-order radiated energy in the CoM frame
$E_{\rm rad}=\hat{P}\cdot P_{\rm rad}$,
produced in collaboration with Plefka and Steinhoff~\cite{Jakobsen:2021lvp},
in which performing the required integrals necessitated a PN expansion ---
now we no longer need to do so.
While knowledge of both $\theta$ and $P^\mu_{\rm rad}$ allows one to
reconstruct $\Delta p_i^\mu$ for aligned spins,
this is not true in general --- thus, in this work we fill in
the missing pieces from \Rcite{Riva:2022fru}.
However, we note that a corresponding expression for $J^\mu_{\rm rad}$
at 3PM order is still lacking (the leading-order 2PM is known, see \Eqn{eq:jRad} below).

The 3PM spin kick takes a similar form as the impulse:
\begin{align}
  \Delta S_{1,\rm cons}^{(3;-)\mu}
  &=
  \frac{m_1^2 m_2^2}{|b|^3}
  \Big[
    d_1^{(-)\mu}
    \frac{\text{arccosh}\gamma}{\sqrt{\gamma^2-1}}
    \!+\!
    \sum_{n=1}^3\!
    \Big(
    \frac{m_1}{m_2}
    \Big)^{\!n-2}
    \!d_{n+1}^{(-)\mu}
    \Big],\nn\\
  \Delta S_{1,\rm rad}^{(3;-)\mu}
  &=
  \frac{m_1^2 m_2^2}{|b|^{3}}{\mathcal I}(v)
  d_5^{(-)\mu}\nn\\
  \Delta S_{1,\rm cons}^{(3;+)\mu}
  &=
  \sum_{n=1}^3
  \frac{\pi m_1^2 m_2^2}{|b|^{3}}
  \Big(
  \frac{m_1}{m_2}\!
  \Big)^{n-2}
  d_{n}^{(+)\mu}\,,\\
  \Delta S_{1,\rm rad}^{(3;+)\mu}
  &=
  \frac{\pi m_1^2 m_2^2}{|b|^{3}}\times\nn\\
  &\Big[
    d_{4}^{(+)\mu} +
    d_{5}^{(+)\mu}
    \frac{\text{arccosh}\gamma}{\sqrt{\gamma^2-1}}
    \!+\!
    d_{6}^{(+)\mu}
    \log\left(\frac{1+\gamma}2\right)\Big],\nn
\end{align}
but in this case $\Delta S_1^{(3;-)\mu}$ is associated with real integrals
and $\Delta S_1^{(3;+)\mu}$ with imaginary integrals~\eqref{eq:integralsReality}.
The vectors $d_n^{(\pm)\mu}$ are given by
\begin{align}
  d_n^{(\pm)\mu}&=\vec{g}_n^{\,\,(\pm)}(\gamma,C_{{\rm E},i})\cdot{\vec{\rho^\prime}_\pm}^{\,\mu}\,,
\end{align}
with basis elements ${\vec{\rho^\prime}_\pm}^{\,\mu}$ even/odd under time reversal:\footnote{This discussion has been corrected from a prior version of this paper;
results in the ancillary file remain unchanged.
}
\begin{subequations}
\begin{align}
  &
  {\vec{\rho^\prime}_+}^{\,\mu}
  =
  \bigg\{
  \frac{a_1\cdot\hat b}{|b|}
  \hat b^\mu
  ,
  \frac{a_1\cdot v_2}{|b|}
  v_i^\mu
  ,
  \frac{a_i\cdot \hat b
    \,
    a_j\cdot \hat L}{|b|^2}
  \hat b^\mu
  ,
  \frac{a_1\cdot \hat b
    \,
    a_i\cdot \hat b}{|b|^2}
  \hat L^\mu
  ,
  \nn
  \\
  &\qquad\qquad
  \frac{a_1\cdot v_2
    \,
    a_i\cdot v_{\bar \imath}}{|b|^2}
  \hat L^\mu
  ,
  \frac{a_i\cdot \hat L
    \,
    a_j\cdot v_{\bar\jmath}}{|b|^2}
  v_k^\mu
  \bigg\}
  \ ,
  \\
  &
  {\vec{\rho^\prime}_-}^{\,\mu}
  =
  \bigg\{
  \frac{a_1\cdot v_2}{|b|}
  \hat b^\mu
  ,
  \frac{a_1\cdot\hat b}{|b|}
  v_i^\mu
  ,
  \frac{a_i\cdot \hat L
    \,
    a_j\cdot v_{\hat\jmath}}{|b|^2}
  \hat b^\mu
  ,
  \frac{
    a_i\cdot \hat b
    \,
    a_j\cdot v_{\hat\jmath}
  }{|b|^2}
  \hat L^\mu
  ,
  \nn
  \\
  &\qquad\qquad
  \frac{a_i\cdot \hat b
    \,
    a_j\cdot \hat L
  }{|b|^2}
  v_k^\mu
  \bigg\}
  \ .
\end{align}
\end{subequations}
Only the basis elements of this list with at least one factor of $a_1^\mu$ are relevant for the kick $\Delta S_1^\mu$.
Again, we refer the interested reader to the ancillary file for fully explicit results.

\subsection{Scattering Angles}

Conservative dynamics with spin vectors aligned to the scattering plane
are described by a single angle:
\begin{align}\label{eq:simplePar}
  \Delta p^\mu_{1,\rm cons}
  = p^\mu(\cos\theta_{\rm cons}-1) + p_{\infty}\hat{b}^\mu\sin\theta_{\rm cons}\,,
\end{align}
where $p^\mu$ is the CoM momentum~\eqref{eq:comMom}.
However, generic spins and radiative effects both require a generalization of this simple parameterization.
Generic spins result in non-planar motion and a non-zero spin kick;
radiative effects in loss of total four-momentum $P_{\rm rad}^\mu$ ---
we will present a more generic parametrization below~\eqref{eq:saketh}.

For generic mis-aligned spin directions the impulse and spin kick
are parameterized in spherical coordinates in terms of several angles.
We focus on the following two, including radiation-reaction effects:
\begin{subequations}
  \label{eq:genericAngles}
  \begin{align}\label{eq:angle1}
    \sin
    \bigg(
    \frac{\theta_1}{2}
    \bigg)
    &=
    \frac{|\Delta p_{1}|}{2\pin}
    \ ,
    &
    \sin
    \bigg(
    \frac{\theta_2}{2}
    \bigg)
    &=
    \frac{|\Delta p_{2}|}{2\pin}
    \ ,
    \\
    \sin(\phi_1)&=
    -\frac{\hat{b}\cdot \Delta p_{1}}{\pin}\,,
    &
    \sin(\phi_2)&=
    \frac{\hat{b}\cdot \Delta p_{2}}{\pin}
    \ .
  \end{align}
\end{subequations}
The conservative counterparts of these angles, $\tc$ and $\pc$,
are defined by instead inserting $\Delta p_{i,\rm cons}^\mu$ on the right-hand side.
In this case the particle label on the angles is superficial as $\Delta p^\mu_{1, \rm cons}=-\Delta p^\mu_{2,\rm cons}$.
For aligned spins at 3PM order these two definitions are equivalent to each other: $\theta_i=\phi_i$,
which using \Eqn{eq:simplePar} holds to all PM orders for strictly conservative scattering.
Up to linear order in spin the angles equate even for mis-aligned spin vectors:
$\theta_i=\phi_i+\cO(S^2)$,
as dependence on the spin vectors $S_i^\mu$ only enters through the spin-orbit terms $L\cdot S_i$.
Surprisingly though, and in contrast to the use of spherical coordinates,
we will find that only one of these angles suffices to fully describe the conservative impulse and spin kick.

In the conservative case the interpretation of $\tc$ and $\pc$ is simple.
First, $\pc$ measures the total scattering angle in the CoM frame in the plane spanned by $\hat b^\mu$ and $\hat p^\mu$.
Second, $\tc$ measures the total angle between the initial and final momentum (which may point out of the initial plane).
Including radiation-reaction we may still compute $\theta_i$ and $\phi_i$ using
Eqs.~\eqref{eq:genericAngles} although their physical interpretation as scattering angles is less clear.
We also note that, while $\theta_i$ is independent of the choice of SSC,
$\phi_i$ is not due to the manifest dependence on $b^\mu$, which transforms under SUSY shifts.
To put it another way: the notion of initial plane of scattering depends on the SSC.
For this reason we will mostly focus on $\tc$ and later use it to parameterize the Hamiltonian in \Sec{sec:ham}.

We expand the angle $\theta_i$ in $G$ and spins:
\begin{align}\label{eq:angleCov2}
  &
  \frac{\theta_k}{\Gamma}
  =
  \sum_{n=1}^3
  \bigg(
  \frac{G M}{|b|}
  \bigg)^n
  \bigg[
    \thii{n;0}
    -
    \sum_i \thii{n;1,i} \frac{\hat L\cdot a_i}{|b|}
    \\
    &\qquad
    +
    \sum_{i,j}
    \frac{a_i^\mu a_j^\nu}{|b|^2}
    \bigg(
    -\thii{n;2,1,i,j} \eta^{\mu\nu}
    +\thii{n;2,2,i,j} \hat b^\mu \hat b^\nu
    \nn
    \\
    &\qquad
    +\thii{n;2,3,i,j} v^\mu_{\bar \imath} v^\nu_{\bar \jmath}
    +\thii{n;2,4,i,j}_k \hat b^\mu v^\nu_{\bar \jmath}
    \bigg)
    \bigg]
  +\cO(S^3,G^4)\,.\nn
\end{align}
Here $i$ and $j$ take the values 1,2.
The coefficients $\thii{n;A}$ are functions only of $\gamma$, $\nu$ and $C_{{\rm E},i}$.
Note that only the final coefficients $\thii{n;2,4,i,j}_k$ depend on the particle label.
The coefficients $\thii{3;A}$ are provided in Appendix~\ref{sec:angles} and full expressions for the angles in the ancillary file.
Expansion coefficients for $\tc$, i.e.~$\thiio{n;A}$, are defined in an equivalent manner.

For aligned spins we verify our results for $\theta_1=\theta_2$ against several results from the literature.
First, we reproduce the result of our earlier work~\cite{Jakobsen:2022fcj} 
where the radiative part of $\theta_i$ was computed using linear response
(see \Sec{sec:linResponse})
and has subsequently been extended to all spin orders
by Alessio and Di Vecchia~\cite{Alessio:2022kwv}.
Second, we match our results for the probe limit and comparable-mass PN results \cite{Damgaard:2022jem,Vines:2018gqi,Antonelli:2020ybz}.
Finally, for mis-aligned spin vectors in the high-energy limit where
we let $\gamma\to\infty$ while keeping $E$ and $a_i^\mu$ constant we recover a finite result:
\begin{widetext}
  \begin{align}
    \begin{aligned}
      &
      \theta_i
      =
      4
      \frac{G E}{|b|}
      \bigg[
        1
        +
        \frac{\hat L\cdot a_+}{|b|}
        -
        \frac{
          2
          a_+^2
          +
          3
          \big(
          \hat b\cdot a_+
          \big)^2
        }{2|b|^2}
        +
        \sum_j
        C_{{\rm E},j}
        \frac{a_j^2+2\big(\hat b\cdot a_j
          \big)^2}{|b|^2}
        \bigg]
      +
      \frac{32}{3}
      \bigg(\frac{G E}{|b|}\bigg)^3
      \bigg[
        1
        +
        3\frac{\hat L\cdot a_+}{|b|}
        \\&\qquad
        -
        \frac{3}{20}
        \frac{
          41 a_+^2
          +
          a_-^2
          +
          50\big(
          \hat b\cdot a_+
          \big)^2}{|b|^2}
        -\frac{945\pi}{8192}C_{{\rm E},i}
        \frac{
          \hat b\cdot a_i
          \,
          v_{\bar \imath} \cdot a_i
        }{|b|^2}
        +
        \frac65 \sum_j C_{{\rm E},j} \frac{
          2 a_j^2
          +
          5\big(
          \hat b\cdot a_j
          \big)^2}{|b|^2}
        \bigg]
      +
      \cO(\gamma^{-1/2},G^4)\,.
    \end{aligned}
  \end{align}
\end{widetext}
Here we use the notation $a^\mu_\pm=a^\mu_1\pm a^\mu_2$.
Cancellations between the conservative and radiative pieces are essential
to ensure the finiteness of this result.
Note the dependence on the particle label in the second term of the second line,
which disappears for Kerr black holes.

Finally, let us discuss parameterizations of the full radiative observables.
One may introduce Lorentz transformations ${\Lambda_i}^\mu_{\ \nu}$ that transform
the initial momenta and spin vectors to the final ones \cite{Bini:2017xzy,Vines:2017hyw,Liu:2021zxr}:
\begin{subequations}
  \label{eq:transformation}
  \begin{align}
    \Delta p_i^\mu &= ({\Lambda_i}^\mu_{\ \nu}- \delta^\mu_\nu) p_i^\nu
    \ ,
    \label{eq:impulse}
    \\
    \Delta S_i^\mu &= ({\Lambda_i}^\mu_{\ \nu}- \delta^\mu_\nu) S_i^\nu
    \ .
  \end{align}
\end{subequations}
The same transformation acts on both $p_i^\mu$ and $S_i^\mu$:
one sees this naturally given the requirement that 
$p_i^2$, $S_i^2$ and $p_i\cdot S_i$ must all be explicitly conserved.
Conservation of $p_i^2$ ($S_i^2$) implies that the final momentum (spin vector) is given by a boost (rotation) of the initial one.
Conservation of $p_i\cdot S_i$ implies that the boost and rotation may be combined into a single Lorentz transformation.

\section{Beyond Linear response}
\label{sec:linResponse}

The Bini-Damour linear response relation is used to
infer linearly radiative contributions to the scattering angle $\theta$
from (angular) momentum loss at lower-PM orders. 
For aligned spins \cite{Bini:2012ji,Damour:2020tta,Bini:2021gat}
\begin{align}\label{eq:linResponse}
  \theta_{\rm rad}
  =-\frac12\left(\frac{\partial\theta}{\partial J}J_{\rm rad}
  +\frac{\partial\theta}{\partial E}E_{\rm rad}\right)\,, 
\end{align}
where $J_{\rm rad}$ and $E_{\rm rad}$ are respectively the
angular momentum and energy losses.
In \Rcite{Jakobsen:2022fcj} this was used to deduce the radiative part
of the quadratic-in-spin 3PM scattering angle 
$\theta_{\rm rad}^{(3)}$ for aligned spins,
which we have now re-confirmed with our full calculation of $\Delta p_1^{(3)\mu}$.
At 3PM order, given that $E_{\rm rad}\sim\cO(G^3)$ the second term plays no role:
the linear response is entirely accounted for by the radiated angular momentum $J_{\rm rad}\sim\cO(G^2)$.

Using our newly derived observables we have checked and can confirm that 
at 3PM order the linear response relation \eqref{eq:linResponse}
generalizes for mis-aligned spin vectors to
\begin{align}\label{eq:linResponseFull}
  \Delta p^{(+)\mu}_{1,\rm rad}=
  \frac12\left(\frac{\partial\Delta p^\mu_1}{\partial J^\nu}J_{\rm rad}^{\nu}+
  \frac{\partial\Delta p^\mu_1}{\partial P^\nu}P_{\rm rad}^{\nu}\right)\,.
\end{align}
Again, the part of this formula carrying $P_{\rm rad}^\mu$ vanishes at 3PM order,
but we include it to maintain the analogy with \Eqn{eq:linResponse};
in \Eqn{eq:parityP} we defined $\Delta p^{(+)\mu}_i$ as the part of the impulse
even under a time-reversal operation, where $v_i^\mu\to-v_i^\mu$.
As a special case of Eq.~\eqref{eq:linResponseFull} we derive a linear response relation for $\theta_i$
\eqref{eq:angle1} at 3PM order:
\begin{align}
  \theta_{i,\rm rad}^{(+)}
  =
  -\frac12
  \left(
  \frac{\partial\theta_i}{\partial J^\mu}J^\mu_{\rm rad}
  +\frac{\partial\theta_i}{\partial P^\mu}P^\mu_{\rm rad}
  \right)\,, 
\end{align}
where $\theta_{i,\rm rad}=\theta_i-\theta_{\rm cons}$
and the $(+)$ superscript was defined in Eq.~\eqref{eq:timeReversal}.

For the spin kick we learn about the odd part $\Delta S^{(-)\mu}_i$:
\begin{align}\label{eq:linResponseFullA}
  \Delta S^{(-)\mu}_{1,\rm rad}=
  \frac12\left(\frac{\partial\Delta S_1^\mu}{\partial J^\nu}J_{\rm rad}^{\nu}+
  \frac{\partial\Delta S_1^\mu}{\partial P^\nu}P_{\rm rad}^{\nu}\right)\,.
\end{align}
The $J^\nu$-derivative is equivalent to an $L^\nu$-derivative~\eqref{eq:jVec}:
when taking these vectorial derivatives, we ignore all constraints (e.g.~$L\cdot p_i=0$) and
treat the vectors involved ($L^\mu$, $p_i^\mu$, $a_i^\mu$) as independent.
All dependence on $b^\mu$ should be re-expressed in terms of $L^\mu$
by inverting $L^\mu=-E^{-1}{\eps^\mu}_{\nu\rho\sigma}b^\nu p_1^\rho p_2^\sigma$.

Both of these linear response relations involve the full vectorial radiated
angular momentum $J_{\rm rad}^\mu$.
We require it only up to 2PM order:
\begin{align}\label{eq:jRad}
  &J_{\rm rad}^{(2)\mu} =\Re\bigg[-\frac{4M^3\nu^2(2\gamma^2-1)}{|b|\Gamma}{\cal I}(v)\zeta^\mu\\[-5pt]
    &\quad\times\bigg(1+\frac{2v\,a_3\cdot \zeta}{|b|(1+v^2)}
    +\frac{(a_3\cdot\zeta)^2}{|b|^2}
    -\sum_{i=1}^2\frac{C_{{\rm E},i}}{|b|^2}(a_i\cdot\zeta)^2
    \bigg)\bigg]\,,\nn
\end{align}
where $v=\sqrt{\gamma^2-1}/\gamma$ is the relative velocity,
$a_3^\mu=a_1^\mu+a_2^\mu$ and the complex vector is $\zeta^\mu=\hat{L}^\mu+i\hat{b}^\mu$;
the universal prefactor ${\cal I}(v)$ was given in \Eqn{eq:univPrefactor}.

For zero or aligned spins,
one may straightforwardly show that the new linear response relation~\eqref{eq:linResponseFull}
reduces to the Bini-Damour formula~\eqref{eq:linResponse} by inserting \cite{Saketh:2021sri}:
\begin{align}\label{eq:saketh}
  \Delta p_1^{\mu}&=p_\infty\hat{b}^\mu\sin\theta-v_2\cdot P_{\rm rad}\frac{\gamma v_1^\mu-v_2^\mu}{\gamma^2-1}
  +(\cos\theta\!-\!1)p^\mu\,,
\end{align}
which holds up to the desired 3PM order.
This schematic form of the impulse shows that
$\Delta p_1^\mu$ is fully characterized by $\theta$ and $P_{\rm rad}^\mu$:
as $\Delta p^{(+)\mu}=p_\infty\sin\theta\,\hat{b}^\mu$
knows only about the scattering angle,
the linear response relationship yields no information concerning $P_{\rm rad}^\mu$.
We use the fact that, for aligned spins, $J_{\rm rad}^\mu=-J_{\rm rad}\hat{L}^\mu$.

However, in this section we generalize beyond simple linear response.
The linear  response relation~\eqref{eq:linResponseFull} forms part of
a more general pair of relationships that allow us to reconstruct conservative and radiative parts of
the scattering observables at higher-PM orders:
\begin{subequations}~\label{eq:respRels}
  \begin{align}
    \Delta p^\mu_{i,\rm cons}=
    \sfrac12\big(&\Delta p_i^\mu(L^\mu,p_i^\mu,S_i^\mu)\label{eq:consRelP}\\
    +&\Delta p_i^\mu(L^\mu+\Delta L^\mu,-p_i^\mu-\Delta p_i^\mu,S_i^\mu+\Delta S_i^\mu)\big),\nn\\
    \Delta p^\mu_{i,\rm rad}=
    \sfrac12\big(&\Delta p_i^\mu(L^\mu,p_i^\mu,S_i^\mu)\label{eq:radRelP}\\
    -&\Delta p_i^\mu(L^\mu+\Delta L^\mu,-p_i^\mu-\Delta p_i^\mu,S_i^\mu+\Delta S_i^\mu)\big),\nn
  \end{align}
\end{subequations}
for the impulse, and
\begin{subequations}\label{eq:kickRel}
  \begin{align}
    \Delta S^\mu_{i,\rm cons}=
    \sfrac12\big(&\Delta S_i^\mu(L^\mu,p_i^\mu,S_i^\mu)\label{eq:consRelA}\\
    -&\Delta S_i^\mu(L^\mu+\Delta L^\mu,-p_i^\mu-\Delta p_i^\mu,S_i^\mu+\Delta S_i^\mu)\big),\nn\\
    \Delta S^\mu_{i,\rm rad}=
    \sfrac12\big(&\Delta S_i^\mu(L^\mu,p_i^\mu,S_i^\mu)\label{eq:radRelA}\\
    +&\Delta S_i^\mu(L^\mu+\Delta L^\mu,-p_i^\mu-\Delta p_i^\mu,S_i^\mu+\Delta S_i^\mu)\big),\nn
  \end{align}
\end{subequations}
for the spin kick;
we will define the split $\Delta X=\Delta X_{\rm cons}+\Delta X_{\rm rad}$
of the full observables into conservative and radiative parts below.
We interpret all observables as real functions of the initial kinematic vectors:
the orbital angular momentum vector $L^\mu$~\eqref{eq:Lvec},
the spin vectors $S_i^\mu=m_ia_i^\mu$, and the momenta $p_i^\mu=m_iv_i^\mu$.

\subsection{Derivation}

We define the conservative part of a single-operator expectation value
as the average of its value evaluated in the in-in and out-out prescriptions:
\begin{equation}
  \braket{\cO}_{\rm cons}:=\sfrac12
  \left(\braket{\cO}_{\rm in-in}+\braket{\cO}_{\rm out-out}\right)\,.
\end{equation}
The expectation
$\braket{\cO}_{\rm out-out}$ is computed using precisely the same Feynman rules
as $\braket{\cO}_{\rm in-in}$, but with advanced propagators pointing
towards the outgoing line instead of retarded,
both on the worldlines and in the bulk.
At 3PM order, one may verify by explicit calculation that
this definition of the conservative dynamics coincides
precisely with evaluating integrals only in the potential region ---
the approach previously taken for the conservative
3PM spinning dynamics in \Rcite{Jakobsen:2022fcj}.

Specializing to the impulse $\Delta p_i^\mu$, we therefore have
\begin{align}\label{eq:responsePrelim}
  &\Delta p^\mu_{i,\rm cons}\\
  &\!\!=\sfrac12\big(\Delta p^\mu_{i,\rm in-in}(L_{-}^\mu,p_{i-}^\mu,S_{i-}^\mu)\!+\!
  \Delta p^\mu_{i,\rm out-out}(L_{+}^\mu,p_{i+}^\mu,S_{i+}^\mu)\big).\nn
\end{align}
While $\Delta p^\mu_{i,\rm in-in}$ is given in terms of background
parameters defined at past infinity ($-$~subscript),
$\Delta p^\mu_{i,\rm out-out}$ is evaluated in terms of
parameters defined at future infinity ($+$ subscript).
As we prefer to express $\Delta p^\mu_{i,\rm cons}$ in terms of initial variables we insert
\begin{subequations}
  \begin{align}
    L_-^\mu&=L^\mu\,, & L_+^\mu&=L^\mu+\Delta L^\mu\,,\\
    p_{i-}^\mu&=p_i^\mu\,, & p_{i+}^\mu&=p_{i}^\mu+\Delta p_i^\mu\,,\\
    S_{i-}^\mu&=S_i^\mu\,, & S_{i+}^\mu&=S_{i}^\mu+\Delta S_i^\mu\,.
  \end{align}
\end{subequations}
Thus $p_{i+}^\mu$ and $S_{i+}^\mu$ are given using our pre-existing
knowledge of the impulse $\Delta p_i^\mu$ and spin kick
$\Delta S_i^\mu$.
We can infer $\Delta L^\mu$ --- and therefore $L_+^\mu$ ---
up to 2PM order from the known 2PM angular momentum loss $J_{\rm rad}^\mu$~\eqref{eq:jRad}.
Using \Eqn{eq:jVec}
\begin{align}
  \Delta J^\mu&=-J_{\rm rad}^\mu\\
  &=\Delta L^\mu+\sum_i\sfrac1{m_i}\big(\hat{P}\cdot\Delta p_i\,S_i^\mu+
  \hat{P}\cdot(p_i+\Delta p_i)\Delta S_i^\mu\nn\\[-5pt]
  &\qquad\qquad\qquad\,\,\,\,
  -\hat{P}\cdot\Delta S_i\,p_i^\mu-\hat{P}\cdot(S_i+\Delta S_i)\Delta p_i^\mu\big),\nn
\end{align}
which we can rearrange to find $\Delta L^\mu$ ---
ignoring the (linear) momentum loss $P_{\rm rad}^\mu\sim\cO(G^3)$.
In the non-spinning case $\Delta L^\mu=-J_{\rm rad}^\mu$,
i.e.~the change in the orbital angular momentum
vector is given precisely by the total loss of angular momentum.

Finally, to obtain \Eqn{eq:respRels} we use the fact that
\begin{align}\label{eq:timeReverse}
  \Delta p^\mu_{i,\rm out-out}(L^\mu,p_i^\mu,S_i^\mu)=
  \Delta p^\mu_{i,\rm in-in}(L^\mu,-p_i^\mu,S_i^\mu)\,,
\end{align}
which simply tells us that, having computed $\Delta p^\mu_{i,\rm in-in}$,
we may easily derive $\Delta p^\mu_{i,\rm out-out}$ by continuing $p_i^\mu\to-p_i^\mu$.
This works because the time-reversal operation induces a change of sign on the $i0$ prescription
of the propagators \eqref{eq:gravProp} and \eqref{eq:wlPropagators}.
For the graviton propagator \eqref{eq:gravProp} ${\rm sgn}(k^0)i0=(k\cdot v_i)i0$:
the sign on the energy component of $k^\mu$ is defined by the direction of either velocity vector $v_i^\mu$.
For the worldline propagators \eqref{eq:wlPropagators}, $\omega\to-\omega$:
the $z_i$ propagator~\eqref{eq:propZ} remains the same, but with $i0\to-i0$,
while the $\psi_i^\prime$ propagator~\eqref{eq:propPsi} also picks up an overall sign.
This overall sign is compensated for in the WQFT Feynman rules for by the vertices,
which are themselves invariant under time reversal except for
$S_i^{\mu\nu}={\eps^{\mu\nu}}_{\rho\sigma}v_i^\rho S_i^\sigma$,
which flips as $S_i^{\mu\nu}\to-S_i^{\mu\nu}$.
Each time we propagate an internal spin mode we pick up a factor of $S_i^{\mu\nu}$,
which compensates for the additional sign.

The derivation of \Eqn{eq:kickRel} for the spin kick proceeds similarly,
although in this case as $\Delta S_i^\mu$ is defined indirectly
via $\Delta p_i^\mu$ and $\Delta S_i^{\mu\nu}$~\eqref{eq:spinKick}
we obtain the different relative signs.
One can see why this is necessary by examining the spin kick at 1PM order:
\begin{align}
  \Delta S_1^{(1)\mu}=\frac{
    4 m_1m_2 a_{1\nu} b^{[\nu}(v_1-2\gamma v_2)^{\mu]}
  }{|b|^2\sqrt{\gamma^2-1}}
  +\cO(S^2)\,,
\end{align}
which is of course purely conservative.
It is odd under the time-reversal $v_i^\mu\to-v_i^\mu$,
which agrees with \Eqn{eq:radRelA}:
we have $\Delta S_i^{(1)\mu}(L^\mu,p_i^\mu,S_i^\mu)=-\Delta S_i^{(1)\mu}(L^\mu,-p_i^\mu,S_i^\mu)$.

\subsection{Interpretation}

We can now derive $\Delta p^{(m;-)\mu}_{i,\rm cons}$ at any PM order $m$.
To do so, we insert the PM decomposition~\eqref{eq:pmDecomp}
into \Eqn{eq:consRelP} and perform a Taylor-series expansion of the right-hand side,
picking out the desired PM order $m$:
\begin{align}
  \Delta p_{i,\rm cons}^{(m)\mu}&=\sfrac12\big(\Delta p_i^{(m)\mu}(L^\mu,p_i^\mu,S_i^\mu)+
  \Delta p_i^{(m)\mu}(L^\mu,-p_i^\mu,S_i^\mu)\nn\\
  &\quad+\frac{\partial\Delta p_i^{(m-1)\mu}(L^\mu,-p_i^\mu,S_i^\mu)}{\partial L^\nu}\Delta L^{(1)\nu}\nn\\
  &\quad+\sum_{j=1}^2\bigg(-\frac{\partial\Delta p_i^{(m-1)\mu}(L^\mu,-p_i^\mu,S_i^\mu)}{\partial p_j^\nu}\Delta p_j^{(1)\nu}\nn\\
  &\quad\qquad\qquad\!\!
  +\frac{\partial\Delta p_i^{(m-1)\mu}(L^\mu,-p_i^\mu,S_i^\mu)}{\partial S_j^\nu}\Delta S_j^{(1)\nu}\bigg)\nn\\
  &\quad+\cdots\big)\,.
\end{align}
Taking the difference between this formula and its counterpart with $p_i^\mu\to-p_i^\mu$
gives $\Delta p^{(m;-)\mu}_{i,\rm cons}$ on the left-hand side,
and the first two terms on the right-hand side cancel out.
Thus, $\Delta p^{(m;-)\mu}_{i,\rm cons}$ is given entirely by lower-PM observables.
Similarly, using \Eqn{eq:consRelA} we may predict $\Delta S^{(m;+)\mu}_{i,\rm cons}$.

Using~\Eqns{eq:radRelP}{eq:radRelA} by the same procedure we may determine
$\Delta p^{(+)\mu}_{i,\rm rad}$ and  $\Delta S^{(-)\mu}_{i,\rm rad}$.
However, at 3PM order an additional simplification is possible:
using the fact that the conservative and radiative observables
have opposite behaviors under $v\to -v$.
As $J_{\rm rad}^\mu$ is the only non-zero radiative observable at 2PM order,
it follows that all other contributions to the linear response relation cancel out at 3PM order,
leaving us with \Eqns{eq:linResponseFull}{eq:linResponseFullA} as proposed earlier.
This we have checked carefully by direct calculation.

We anticipate that \Eqn{eq:kickRel} will be useful for future 4PM computations,
as we will not need to calculate the complete radiative observables:
for the impulse we need only calculate $\Delta p_{1,\rm cons}^{(4;+)\mu}$
and $\Delta p_{1,\rm rad}^{(4;-)\mu}$ directly.
This cuts down on the regions within which we need to evaluate the master integrals:
only the conservative sector for the real integrals,
and the radiative sector for the pseudoreal integrals.
However, in the radiative sector this is predicated on our knowing the 3PM angular momentum
loss $J_{\rm rad}^{(3)\mu}$, which currently we have only up to the leading 2PM order~\eqref{eq:jRad}
in the spinning case.
For non-spinning bodies, the 3PM angular momentum loss has been determined
and used to infer contributions to 4PM scattering observables~\cite{Manohar:2022dea};
a similar concept will certainly apply in the presence of spin.

\section{Hamiltonian}\label{sec:ham}

Let us now focus on the strictly conservative part of the dynamics,
encoded by $\Delta p_{i,\rm cons}^\mu$ and $\Delta S_{i,\rm cons}^\mu$.
Computing a 3PM quadratic-in-spin Hamiltonian maps these unbound observables into bound dynamics,
which in the spinning context is especially useful given the current lack of direct analytic continuations
between bound and unbound observables with generic mis-aligned spin directions.
In line with recent literature on Post-Minkowskian dynamics~\cite{Bern:2020buy,FebresCordero:2022jts,Bern:2022kto}
we work in the CoM-frame with canonical variables $\vct{p}(t)$ and $\vct{x}(t)$
describing the relative momentum and position of the two bodies respectively.
These dynamical variables satisfy canonical Poisson brackets:
\begin{subequations}
  \begin{align}
    &
    \{\vct{x}^m(t),\vct{p}^n(t)\}_{\rm P.B.}=\delta^{mn}
    \ ,
    \\
    &
    \{\vct{S}^m_i(t),\vct{S}^n_i(t)\}_{\rm P.B.}=\epsilon^{mnk}\vct{S}_i^k(t)
    \ .
  \end{align}
\end{subequations}
The spin of each body is described by the
spin vectors $\vct{S}_i(t)$, with the canonical SSC described in Sec.~\ref{sec:canonical}.

The Hamiltonian $H$ is fixed in isotropic gauge,
meaning that it does not depend on $\vct{x}(t)\cdot\vct{p}(t)$.
It takes the general form
\begin{align}\label{eq:hamiltonian}
  H
  \big(
  \vct{x},\vct{p},\vct{S}_i
  \big)
  \!=\!
  \sqrt{\vct{p}^2+m_1^2}\!+\!\sqrt{\vct{p}^2+m_2^2}
  \!+\!
  V
  \big(
  \vct{x},\vct{p},\vct{S}_i
  \big)
\end{align}
with gravitational potential
\begin{align}
  &V
  \big(
  \vct{x},\vct{p},\vct{S}_i
  \big)
  =
  \sum_A
  \cO^A\,
  V^A(\vct{x},\vct{p})
  +\cO(S^3)\\
  &=V^{(0)}
  \!\!
  +
  \!\!
  \sum_i V^{(1,i)} \cO^{(1,i)}
  \!
  +
  \!\!
  \sum_{i,j,a}
  V^{(2,a,i,j)} \cO^{(2,a,i,j)}
  \!
  +
  \!
  \cO(S^3).
  \nn
\end{align}
The potential is expanded in spin structures:
\begin{subequations}
  \begin{align}
    \cO^{(0)}
    &=1\,,
    \\
    \cO^{(1,i)}
    &=
    \frac{
      ({\vct{x}}\times{\vct{p}})
      \cdot{\vct{a}}_i
    }{
      |{\vct{x}}|^2
    }\,,
    \\
    \cO^{(2,1,i,j)}
    &=
    \frac{
      {\vct{a}}_i\cdot{\vct{a}}_j
    }{
      |{\vct{x}}|^2
    }\,,
    \\
    \cO^{(2,2,i,j)}
    &=
    \frac{
      {\vct{x}}\cdot{\vct{a}}_i
      {\vct{x}}\cdot{\vct{a}}_j
    }{
      |{\vct{x}}|^4
    }\,,
    \\
    \cO^{(2,3,i,j)}
    &=
    \frac{
      {\vct{p}}\cdot{\vct{a}}_i
      {\vct{p}}\cdot{\vct{a}}_j
    }{
      |{\vct{x}}|^2
    }\,,
  \end{align}
\end{subequations}
where ${\mathbf a}_i={\mathbf S}_i/m_i$.
In each case the first index counts the spin order;
subsequent indices count the specific structures involved.
Note that the symmetric spin structures $\cO^{(2,a,1,2)}=\cO^{(2,a,2,1)}$
are counted twice and their coefficients are equal.
Finally, we PM-expand each component:
\begin{align}
  V^A(\vct{x},\vct{p})
  =
  \sum_n
  \left(
  \frac{GM}{|\vct{x}|}
  \right)^n
  \cii{n;A}(\vct{p}^2)\,.
\end{align}
These coefficients $\cii{n;A}$ fully encode the Hamiltonian.

We fix the coefficients $\cii{n;A}(\vct{p}^2)$ by matching observables
computed from the Hamiltonian $H$ with scattering observables from the WQFT.
Hamilton's equations for the dynamical variables are
\begin{align}\label{eq:hamEqns}
  \dot{\vct{x}}
  &=
  \frac{
    \partial H
  }{\partial \vct{p}}\,, &
  \dot{\vct{p}}
  &=
  -
  \frac{
    \partial H
  }{\partial \vct{x}}\,, &
  \dot{\vct{S}}_i
  &=
  -\vct{S}_i
  \times
  \frac{
    \partial H
  }{\partial \vct{S}_i}\,,
\end{align}
and we solve them perturbatively up to third order in $G$:
\begin{subequations}\label{eq:expansions}
  \begin{align}
    \vct{x}(t)&=\vct{x}^{(0)}+\sum_{n=1}^3G^n\vct{x}^{(n)}(t)+\cO(G^4)\,,\\
    \vct{p}(t)&=\vct{p}^{(0)}+\sum_{n=1}^3G^n\vct{p}^{(n)}(t)+\cO(G^4)\,,\\
    \vct{S}_i(t)&=\vct{S}_i^{(0)}+\sum_{n=1}^3G^n\vct{S}_i^{(n)}(t)+\cO(G^4)\,.
  \end{align}
\end{subequations}
The zeroth-order scattering trajectories are
\begin{align}
  \vct{x}^{(0)}
  &=t \frac{\vct{p}_\infty}{\xi E} - \vct{b}_{\rm can}\,, &
  \vct{p}^{(0)} &= \vct{p}_\infty \,, &
  \vct{S}_i^{(0)} = \vct{S}_{i,\infty}\,,
\end{align}
where $p^\mu=(0,\vct{p}_\infty)$, $S^\mu_{i,\rm can}=(0,\vct{S}_{i,\infty})$
and $b^\mu_{\rm can}=(0,\vct{b}_{\rm can})$.
The dimensionless parameter $\xi$ is defined as $\xi=E_1 E_2/E^2$,
where $E=E_1+E_2$ and $E_i=\sqrt{\mathbf{p}_\infty^2+m_i^2}$.
Inserting the expansions of the dynamical variables~\eqref{eq:expansions}
into Hamilton's equations~\eqref{eq:hamEqns} we get perturbative equations of motion at each PM order.
The spatial components of the impulse and spin kick in the CoM frame are then given by
\begin{subequations}
  \begin{align}
    \Delta\mathbf{p}^{(n)}
    &=
    \int_{-\infty}^{\infty} \!\d t\,\,
    \dot{\vct{p}}^{(n)}(t)\,,
    \\
    \Delta\mathbf{S}^{(n)}_{i}
    &=
    \int_{-\infty}^{\infty} \!\d t\,\,
    \dot{\vct{S}}_i^{(n)}(t)\,,
  \end{align}
\end{subequations}
where these are the conservative 3-vector components of $\Delta p_1^\mu$ and $\Delta S_{i,{\rm can}}^\mu$.
The change in the canonical spin vector is given in terms of the covariant spin kick by
\begin{align}
  &\Delta S_{i,\rm can}^\mu-\Delta S_i^\mu\\
  &\qquad=-\frac{m_i\hat{P}\cdot \Delta S_i(\hat{P}^\mu+v_i^\mu)+
    \hat{P}\cdot(S_i+\Delta S_i)\Delta p_i^\mu}{E_i+m_i}\,,\nn
\end{align}
having used the definition of $S^\mu_{i,\rm can}$~\eqref{eq:canonical},
and assuming conservative scattering.

Quite remarkably though, we find that knowledge of $\tc$~\eqref{eq:genericAngles} suffices
in order to fully fix all coefficients in the Hamiltonian.
The impulse and spin kick may then be expressed in terms of the coefficients $c^{(n;A)}(\vct{p}^2)$ and derivatives thereof.
The derivatives come about when we evaluate $\dot{\vct{x}}$ by differentiating $H$ with respect to $\vct{p}$:
\begin{align}
  \frac{\partial{c^{(n;A)}(\vct{p}^2)}}{\partial \vct{p}}
  =
  2\vct{p}
  \frac{\partial{c^{(n;A)}(\vct{p}^2)}}{\partial \vct{p}^2}=
  2\vct{p}\,{c^{(n;A;1)}(\vct{p}^2)}\,,
\end{align}
where $c^{(n;A;m)}:=\partial^m c^{(n;A)}/(\partial \vct{p}^2)^m$ ---
the observables are written as functions of $c^{(n;A;m)}$.
We then match those expressions to the explicit WQFT-derived results and solve for the coefficients.
We print the results until linear in spins here, and the results quadratic in spins in Appendix~\ref{sec:HamCoe}:
\begin{widetext}
  \begin{align}\label{eq:HamCoe}
    \cii{3;0}(\mathbf{p}^2)
    &=
    \left.
    -\frac{\pin^2}{4E\xi}
    \thiic{3;0}
    +
    \frac1{2\pi\pin^2}
    \cD\Big[
      \frac{\pin^4}{E\xi}
      \thiic{1;0}
      \thiic{2;0}
      \Big]
    -
    \frac1{48\pin^2}
    \cD^2\Big[
      \frac{\pin^4}{E\xi}
      (\thii{1;0})^3
      \Big]
    \right|_{\pin\to|\mathbf{p}|}
    \ ,
    \\
    \cii{3;1,i}(\mathbf{p}^2)
    &=\left.
    -\frac{\pin}{4E\xi}
    \thiic{3;1,i}
    +
    \frac1{2\pi\pin^4}
    \cD\Big[\frac{\pin^5}{E\xi}
      \big(
      \thiic{1;0}
      \thiic{2;1,i}
      +
      \thiic{2;0}
      \thiic{1;1,i}
      \big)
      \Big]
    -
    \frac1{16\pin^4}
    \cD^2\Big[
      \frac{\pin^5}{E\xi}
      (\thiic{1;0})^2
      \thiic{1;1,i}
      \Big]
    \right|_{\pin\to|\mathbf{p}|}
    \ .\nn
  \end{align}
\end{widetext}
Here $\phiic{n;A}$ are canonical expansion coefficients defined in Appendix~\ref{sec:angleCan} and related to the covariant expansion coefficients of Eq.~\eqref{eq:angleCov2} by:
\begin{subequations}
  \begin{align}
    \thiic{n;0}
    &=
    \Gamma\,
    \thiio{n;0}
    \\
    \thiic{n;1,i}
    &=
    \Gamma\,
    \Big(
    \thiio{n;1,i}
    +
    n
    \frac{\pin}{E_i+m_i}
    \thiio{n;0}
    \Big)
  \end{align}
\end{subequations}
A particular subtlety here is that $\thiic{n;A}$ is given in terms of the previously-defined
background variables $p_{\infty}$, $\gamma$, $E_i=\sqrt{{\mathbf p}_\infty^2+m_i^2}$ evaluated at past infinity, e.g.
\begin{align}
  \gamma=\frac{p_1\cdot p_2}{m_1m_2}=\frac{E_1E_2+\mathbf{p}_\infty^2}{m_1m_2}\,,
\end{align}
instead of the dynamical momentum $\mathbf{p}(t)$:
we interpolate to the full dynamical coefficients $c^{(n;A;m)}$ simply by replacing one with the other.
We have also introduced the differential operator
\begin{align}
  \cD[X]:=  \frac{\partial  (\pin X)}{\partial \pin}\,,
\end{align}
and $\cD^2[X]=\cD[\cD[X]]$.
The angle and its coefficients are given in Appendix~\ref{sec:angles},
together with full expressions for the Hamiltonian coefficients given in the
ancillary file attached to the \texttt{arXiv} submission of this paper.

We have checked this Hamiltonian numerically against the recent results obtained in Ref.~\cite{FebresCordero:2022jts},
which also included 3PM quadratic-in-spin terms.
Our results complement those by adding $S_1 S_2$ contributions to the Hamiltonian 
together with finite-size effects ($C_{{\rm E},i}$ terms) in the $S_i^2$ sector.
We have also verified that the PN-expansion of this Hamiltonian correctly
reproduces 4PN results in the isotropic gauge~\cite{Levi:2016ofk}.
We did so by PN-expanding the  $c^{(n;A)}(\mathbf{p}^2)$ coefficients in powers of $\mathbf{p}^2$.

Finally, let us observe that: having expressed the coefficients of
the Hamiltonian in terms only of the scattering angle $\phi_{\rm can}$,
this implies that the full conservative scattering observables $\Delta p_{i,\rm cons}^{(3)\mu}$
and $\Delta S_{i,\rm cons}^{(3)\mu}$ may themselves be expressed in terms of this angle.
We would obtain the precise relationship by solving Hamilton's equations again for the impulse and
spin kick, but this time plugging in expressions in terms of $\pc$.
In contrast to the Hamiltonian, the relations thus obtained are gauge-invariant and will be an intriguing topic of future studies.

\section{Unbound-to-bound mappings}\label{sec:binding}

Let us now discuss how our results may be applied to describe bound orbits,
which the now-complete 3PM quadratic-in-spin Hamiltonian~\eqref{eq:hamiltonian} gives us partial access to.
This will allow us to determine the binding energy,
which together with the radiative fluxes may be used to inform complete gravitational waveform models.
In this section we specialize to spin vectors aligned with the orbital angular momentum vector:
\begin{align}
  S_i^\mu=m_ia_i^\mu =
  G m_i^2 \chi_i
  \hat L^\mu\,,
\end{align}
where $\chi_i$ are the directed spin lengths and $S^\mu_{i,\rm can}=S^\mu_i$.
For Kerr black holes $m_i|\chi_i|$ are the radii of the ring singularities, and $-1<\chi_i<1$.
Using \Eqn{eq:lShift} we see that for aligned spins $\hat L^\mu = \hat L_{\rm can}^\mu$;
however, their magnitudes differ and so using \Eqn{eq:lShift} we introduce
\begin{align}
  \lambda &= \frac{|L_{\rm can}|}{G M \mu}=
  \frac{|b|\pin}{GM\mu}
  +
  \frac{
    \cE
  }{
    2
  }
  \Big(
  \chi_+
  +
  \frac{\delta}{\Gamma}
  \chi_-
  \Big)\,,
\end{align}
where $\cE=(E-M)/\mu$ is the reduced binding energy, $\delta=(m_2-m_1)/M$ and
\begin{align}
  \chi_\pm &= \frac{m_1\chi_1\pm m_2\chi_2}{M}\,,\\
  \chi^2_{{\rm E},\pm}&=\frac{C_{{\rm E},1}m_1^2\chi_1^2\pm C_{{\rm E},2}m_2^2\chi_2^2}{M^2}\,.
\end{align}
Using axial symmetry the Hamiltonian $E=H(r,p_r,\lambda,\chi_i)$ depends ---
besides the masses $m_i$ and finite-size coefficients $C_{{\rm E},i}$ ---
on the radial coordinate $r$ and momentum $p_r$, where in axial coordinates
\begin{align}\label{eq:vecP}
  \mathbf{p}^2=p_r^2+\frac{p_\phi^2}{r^2}\,.
\end{align}
The axial momentum $p_\phi=|L_{\rm can}|$ is a constant of motion.
In all of these expressions we leave the dependence on masses $m_i$
and finite-size coefficients $C_{{\rm E},i}$ implicit.

\subsection{Numerical PM binding energy}

\begin{figure}[t!]
  \includegraphics[width=8.5cm]{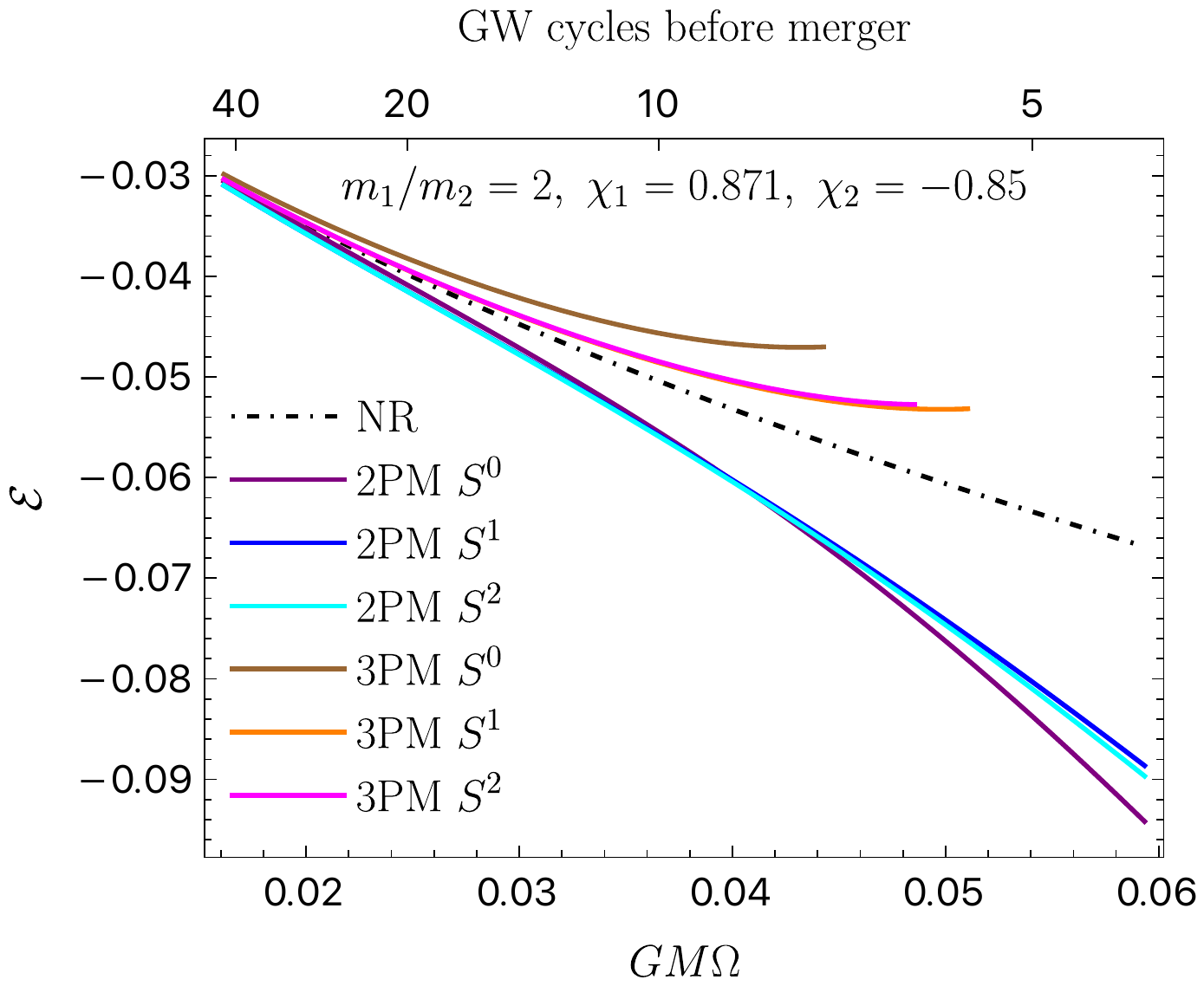}
  \centering
  \caption{\small
    The reduced binding energy $\cE=(H-M)/\mu$ for circular orbits
    determined numerically and plotted as a function of orbital frequency $GM\Omega$
    up to the innermost stable circular orbit.
    It is compared for different PM and spin orders with a Numerical Relativity (NR) simulation
    provided by the SXS collaboration~\cite{Boyle:2019kee}.}
  \label{fig:bindingEnergy}
\end{figure}

In \Fig{fig:bindingEnergy} we plot the reduced binding energy $\cE=(H-M)/\mu$
for circular orbits as a function of the orbital frequency $GM\Omega$ leading up to merger.
It is compared with a Numerical Relativity (NR) simulation provided by the
Simulating eXtreme Spacetimes (SXS) collaboration~\cite{Boyle:2019kee},
extracted in \Rcite{Ossokine:2017dge}.
Our plots are also determined numerically:
within the Hamiltonian we set $p_r=0$ and, using $\dot{p}_r=0=-\partial H/\partial r$,
we solve for $\lambda(r)$ for different orbital separations $r$
and with specific values of $\chi_1$, $\chi_2$.
The reduced binding energy $\cE$ is plotted against the orbital frequency:
\begin{align}\label{eq:frequency}
  x^{3/2}=G M \Omega = \frac{\d\cE}{\d\lambda}\,,
\end{align}
with the number of orbits leading up to merger provided by NR ---
see \Rcites{Antonelli:2019ytb,Khalil:2022ylj} for more details.

The conclusion of these plots is somewhat disappointing:
the quadratic-in-spin part of the Hamiltonian yields little improvement over the spin-orbit contribution.
However, there is a far more noticeable improvement when going from 2PM to 3PM order,
which suggests that producing a 4PM spinning Hamiltonian will be a worthwhile endeavor.
A similar improvement of the 4PM (hyperbolic) Hamiltonian over the 3PM seen in the non-spinning case~\cite{Khalil:2022ylj}
also encourages us in this direction;
however, it will also be important to resum in the test-body limit by feeding these results into a suitable EOB model
\cite{Buonanno:1998gg}.
It is also worth noting that, as these plots are generated for circular orbits,
they do not showcase PM results in the best possible light:
it is anticipated that PM-based results will perform better for highly elliptical orbits,
where the velocity at closest approach between the massive bodies is large~\cite{Khalil:2022ylj}.

\subsection{Analytic PN binding energy}

Working in the PN expansion we may derive precise analytic formulae for the binding energy $\cE$ and periastron advance $\Delta\phi$.
Following closely the discussion in Ref.~\cite{Antonelli:2020ybz}
(see also Refs.~\cite{Kalin:2019inp,Liu:2021zxr})
our starting point is the radial action for unbound orbits:
\begin{align}\label{eq:radAction}
  w_r(\cE,\lambda,\chi_i)=\frac1{GM\mu\pi}\,{\rm Pf}\!\int_{r_{\rm min}}^\infty\!\d r\,\,p_r(r,\cE,\lambda,\chi_i)\,,
\end{align}
where Pf denotes the \emph{partie finie} of the radial action;
the energy constraint $E=H(r,p_r,\lambda,\chi_i)$ can be solved for the radial momentum $p_r$ ---
but we refrain from doing so explicitly.
The innermost point $r_{\rm min}$ is given by the root of $p_r(r_{\rm min},\cE,\lambda,\chi_i)=0$.
This unbound radial action is related to the bound radial action $i_r$ by analytic continuation:
\begin{align}\label{eq:radial}
  i_r(\cE,\lambda,\chi_i)
  =
  w_r(\cE,\lambda,\chi_i)
  -
  w_r(\cE,-\lambda;-\chi_i)\,,
\end{align}
which sends $\lambda\to-\lambda$ and $\chi_i\to-\chi_i$.
For unbound dynamics the reduced binding energy
$\cE>0$ ($\gamma>1$), but now we consider values $\cE<0$ ($0<\gamma<1$)
as we see in \Fig{fig:bindingEnergy}.
The reduced binding energy $\cE$ for circular orbits is derived by setting the bound radial action to zero:
\begin{equation}\label{eq:circular}
  i_r(\cE,\lambda,\chi_i) = 0\,.
\end{equation}
Using Eqs.~\eqref{eq:circular} and~\eqref{eq:frequency} we may write $\cE$ as a
function of $x$, the spins $\chi_\pm$, $\chi_{{\rm E},\pm}$, and $\nu$.

Rather than the Hamiltonian,
we prefer to derive the radial action --- and thus the binding energy $\cE$ ---
from the gauge-invariant conservative scattering angle $\theta_{\rm cons}$.
It is given by a $\lambda$-derivative of the unbound radial action $w_r$:
\begin{align}\label{eq:lDeriv}
  2\pi \frac{\partial}{\partial \lambda} w_r(\cE,\lambda,\chi_i)
  =
  -
  (\theta_{\rm cons}(\cE,\lambda,\chi_i)+\pi)\,,
\end{align}
and by analytic continuation it is related to the periastron advance for bound orbits
\cite{Kalin:2019rwq,Kalin:2019inp,Saketh:2021sri,Cho:2021arx}:
\begin{align}
  \Delta\phi(\cE,\lambda,\chi_i)=\theta_{\rm cons}(\cE,\lambda,\chi_i)+\theta_{\rm cons}(\cE,-\lambda,-\chi_i)\,.
\end{align}
We PM-expand the angle in $\lambda$ as
\begin{align}
  \theta_{\rm cons}=\sum_n\frac{\tilde\theta^{(n)}}{\lambda^n}\,,
\end{align}
where $\tilde\theta^{(n)}$ depends on $\chi_i$ through the ratios $\chi_i/\lambda$.
The analytic continuation in $\lambda$~\eqref{eq:radial} is trivial for all terms in $w_r$ except the
one coming from the non-spinning part of $\tilde\theta^{(1)}$, wherein the dependence on $\lambda$ is $\log(\lambda)$.
With this exception the odd-in-$G$ terms in the PM expansion disappear in the difference~\eqref{eq:radial},
and the analytic continuation of $\log(-\lambda)-\log(\lambda)$ leaves behind a finite piece.
We therefore have
\begin{align}
  i_r(\cE,\lambda,\chi_i)
  =
  -\lambda
  +
  \frac{2\gamma^2-1}{\sqrt{1-\gamma^2}}
  -
  \frac{1}{\pi}
  \sum_n
  \int\!\d\lambda
  \frac{\tilde\theta^{(2n)}}{\lambda^{2n}}\,.
\end{align}
Here $\gamma$ should be re-expressed in terms of $\cE$.
The integral on $\lambda$ is elementary but left unresolved because
of the remaining $\lambda$-dependence inside $\tilde\theta^{(2n)}$.

Naively this result indicates that our 3PM results have no relevance for the mapping to bound results.
However, this obstacle is conveniently circumvented in the PN expansion by use of the so-called impetus formula with PM-coefficients $f_k$:
\begin{align}
  \mathbf{p}^2=p_\infty^2+\sum_{k=1}^\infty\frac{G^k}{r^k}f_k\,.
\end{align}
This is fed into the definition of the scattering angle by way of the radial action~\eqref{eq:radAction}:
\begin{align}
  &\pi+\theta_{\rm cons}
  =-
  \frac{2}{GM\mu}
  \int_{r_{\rm min}}^\infty\d r\,\frac{\partial}{\partial\lambda}p_r(r,\cE,\lambda)
  \\
  &=-2\int_{r_{\rm min}}^\infty\d r\,\frac{\partial}{\partial\lambda}
  \sqrt{\frac1{G^2M^2\mu^2}\left(p_\infty^2+\sum_{k=1}^\infty\frac{G^k}{r^k}f_k\right) -\frac{\lambda^2}{r^2}
  }\,.\nn
\end{align}
This integral has been performed up to high orders in $G$ in \Rcite{Bjerrum-Bohr:2019kec}
and the scattering angle is then expressed in terms of $f_k$.
Knowledge of the 1PM, 2PM and 3PM scattering angles suffices in order to fully determine
$f_{k\leq 3}$, which may in turn be used to reconstruct the leading-PN terms of the angle
at higher PM orders.

In order to reconstruct the quadratic-in-spin binding energy
up to 4PN order, we find it necessary to reconstruct the leading-PN and sub-leading-PN parts of 
$\tilde\theta^{(6)}$ and $\tilde\theta^{(4)}$ respectively,
at both linear and quadratic order in spin.
Following this procedure, we discover that the binding energy is
\begin{widetext}
  \begin{align}\label{eq:bindingResult}
    -2\cE
    =
    x
    &\bigg[
      1
      -
      x
      \frac{9+\nu}{12}
      -
      x^2
      \frac{81-57\nu+\nu^2}{24}
      +\cdots
      \bigg]
    \\
    +
    x^{5/2}
    &\bigg[
      \frac{
        7
        \chi_+
        -
        \delta \chi_-
      }{3}
      +
      x
      \frac{
        (99-61\nu)
        \chi_+
        -
        (45-\nu)
        \delta \chi_-
      }{18}
      +
      x^2
      \frac{
        (405-1101\nu+29\nu^2)\chi_+
        -
        (243-165\nu-\nu^2)\delta\chi_-
      }{24}
      +\cdots
      \bigg]
    \nn
    \\
    -x^3&\bigg[
      \chi_+^2
      +
      \frac{5x}{36}
      \Big(
      (5-6\nu)\chi_+^2
      -44\chi_+\chi_-
      -(1+8\nu)\chi_-^2
      \Big)
      \nn
      \\
      &\quad
      +
      \frac{7x^2}{216}
      \Big(
      (198-680\nu+3\nu^2)\chi_+^2
      -2(171-137\nu)\delta\chi_-\chi_+
      +
      (63-251\nu+56\nu^2)
      \chi_-^2
      \Big)
      \nn
      \\
      &\quad
      +
      \chi_{{\rm E},+}^2
      +\frac{5x}{6}\Big(
      (5-\nu)
      \chi_{{\rm E},+}^2
      -
      2\delta\chi_{{\rm E},-}^2
      \Big)
      +
      \frac{x^2}{72}
      \Big(
      (1125-1025\nu+7\nu^2)\chi_{{\rm E},+}^2
      -
      2\delta(279-70\nu)\chi_{{\rm E},-}^2
      \Big)
      +
      \cdots
      \bigg]\,.
    \nn
  \end{align}
\end{widetext}
At each order in spin, we give the terms up to and including next-to-next-to-leading order in the PN expansion.
With the non-spinning terms appearing up to 2PN order,
the spin-orbit and spin-spin terms appear up to 3.5PN and 4PN order respectively.

\subsection{Radiated energy}

Finally, we may also determine the energy radiated per orbit
in the CoM frame~\cite{Cho:2021arx}:
\begin{align}
  E^{\rm bound}_{\rm rad}(\cE,\lambda,\chi_i)=E_{\rm rad}(\cE,\lambda,\chi_i)-E_{\rm rad}(\cE,-\lambda,-\chi_i)\,,
\end{align}
where $E_{\rm rad}=\hat{P}\cdot P_{\rm rad}$ derives from the full radiative momentum impulse $\Delta p_1^\mu$~\eqref{eq:radP}.
The same analysis was also done successfully by Riva, Vernizzi and Wong~\cite{Riva:2022fru},
who confirmed that this result for the radiated energy for bound orbits
agrees with the corresponding 4PN terms from \Rcite{Cho:2022syn}.
To leading-PN at each spin order, we find that
\begin{align}
  E_{\rm rad}^{\rm bound}=\frac{2\pi\mu^2}{M}\bigg(\frac{148\cE^2}{15\lambda^3}-
  \frac{4(67\chi_+-2\delta\chi_-)\cE^3}{5\lambda^4}&\\
  +\frac{(236(\chi_+^2-\chi_{\rm E,+}^2)+9\chi_-^2)\cE^3}{5\lambda^5}&\bigg)+\cdots\,.\nn
\end{align}
Like in the binding energy~\eqref{eq:bindingResult},
we see that the linear-in-spin terms appear at 1.5PN above the non-spinning;
the quadratic-in-spin terms at 2PN above.

However, we emphasize that, unlike in the case of the binding energy,
this result for the radiated energy does not encode the full leading-PN result.
To see why, we recall that in the non-spinning case the leading-PN radiated energy
may be derived from Einstein's quadrupole formula:
\begin{align}
  E_{\rm rad}^{\rm bound}=\frac{2\pi\mu^2}{M}
  \left(\frac{148\cE^2}{15\lambda^3}+\frac{244\cE}{5\lambda^5}+\frac{85}{3\lambda^7}\right)+\cdots\,.
\end{align}
While we do reproduce the $\lambda^{-3}$ contribution,
gaining access to the $\lambda^{-5}$ and $\lambda^{-7}$ contributions via PM-based scattering calculations
would necessitate 5PM and 7PM calculations respectively: a seemingly impossible task!
However, a more encouraging conclusion was reached in \Rcite{Saketh:2021sri}:
that by instead using the PM-scattering data to fix the form of the (gauge-dependent)
instantaneous fluxes of energy and angular momentum, the leading-PN information could instead be
extracted from a future 4PM derivation.
Such an approach might be especially beneficial in the spinning case,
given the unbound-to-bound mapping's current limitation to aligned spin vectors.
This would be similar to our present use of the conservative scattering observables to reconstruct a Hamiltonian,
a prospect that we leave for future work.

\section{Conclusions}\label{sec:conclusions}

In this paper we have for the first time provided complete expressions for the impulse $\Delta p_1^\mu$ and spin kick $\Delta S_1^\mu$ at third Post-Minkowskian (3PM) order to quadratic order in the spins of two scattering bodies,
including finite-size corrections.
The computation relied on our use of the spinning, supersymmetric WQFT formalism~\cite{Jakobsen:2021lvp,Jakobsen:2021zvh}
and its extension to utilize the Schwinger-Keldysh in-in formalism
\cite{Schwinger:1960qe,Keldysh:1964ud,Jordan:1986ug,Weinberg:2005vy,Galley:2009px,Jakobsen:2022psy}.
These results upgrade the previously obtained conservative observables provided by the present authors~\cite{Jakobsen:2022fcj},
and include knowledge of the total radiated four-momentum $P_{\rm rad}^\mu$
which has also recently been computed using worldline EFT methods~\cite{Riva:2022fru}.
We also wrote down a scattering angle that encapsulates the motion
for arbitrarily mis-aligned spin directions.

Next, we demonstrated how both conservative and radiative parts of these observables
may be reconstructed using an extension of Bini and Damour's linear response relation
\cite{Bini:2012ji,Damour:2020tta,Bini:2021gat},
incorporating the full 2PM radiated angular momentum $J_{\rm rad}^\mu$.
These relations build on a split into conservative and radiative parts as an
average of the full in-in and out-out observables.
It will be exciting to see how these relations may help produce results at 4PM order.
For spin effects this will require a 3PM computation of $J_{\rm rad}^\mu$ with spin,
but one may already explore non-spinning applications of the formula.
Similar studies have already been initiated in the non-spinning case in \Rcite{Manohar:2022dea}.

Using the conservative parts of our results ---
already known from \Rcite{Jakobsen:2022fcj} ---
we constructed a two-body Hamiltonian mapping our unbound results to bound motion.
This Hamiltonian describes the conservative two-body dynamics up to 3PM order
and to quadratic order in their spins --- it complements the Hamiltonian of \Rcite{FebresCordero:2022jts}
by adding the terms $\cO(S_1 S_2)$ and finite-size effects.
We note that the coefficients of the Hamiltonian are more complicated than the scattering observables.
While the observables can easily be reduced to a number of polynomials in $\gamma$,
the Hamiltonian coefficients depend on the center-of-mass variables in a complicated manner.
This is partly due to its gauge dependence,
but also the necessity of using a canonical spin-supplementary condition (SSC).
This canonical Pryce-Newton-Wigner SSC \cite{Pryce:1935ibt,Pryce:1948pf,Newton:1949cq}
we showed is related to the covariant SSC by a supersymmetry transformation.

It is an interesting study to explore whether gauge choices other than the
isotropic gauge could lead to simpler coefficients in the Hamiltonian.
Quite intriguingly, we found it possible to fix all coefficients of the Hamiltonian
by matching to a single scattering angle defined for generic spins.
This in turn leads to the exciting result that the conservative dynamics
for generic spins can be described by a single scalar.
The expressions for the impulse and spin kick in terms of the scattering angle thus obtained are gauge invariant.
They deserve further study and preferably a direct relation highlighting the gauge invariance.
Such relations are similar to the eikonal relations explored in Ref.~\cite{Bern:2020buy}.

We also studied mappings to bound orbits.
This began with using the complete 3PM Hamiltonian to produce
numerical plots of the binding energy for circular orbits leading up to merger,
in comparison with Numerical Relativity (NR) simulations~\cite{Boyle:2019kee}.
We successfully reproduced the known 4PN quadratic-in-spin binding energy,
and the leading-PM radiated energy for bound orbits.
Unfortunately, these plots did not show a significant improvement of the quadratic-in-spin
Hamiltonian over its spin-orbit counterpart;
however, the effect of going from 2PM to 3PM order was more significant.
This calls for the future determination of the 4PM spinning Hamiltonian,
which --- similar to the non-spinning case, and due to the presence of tails ---
encounters non-localities that distinguish between bound and unbound dynamics
\cite{Dlapa:2021npj,Dlapa:2021vgp,Dlapa:2022lmu,Bern:2021dqo,Bern:2021yeh,Khalil:2022ylj}.
To inform realistic waveform models, it will also be important to incorporate knowledge of the test-body
limit by way of the effective-one-body (EOB) formalism~\cite{Buonanno:1998gg,Damour:2008yg}.

Finally, we determined the energy radiated per orbit in the CoM frame from an
appropriate unbound-to-bound mapping~\cite{Cho:2021arx}.
Current limitations in this mapping restrict us to considering only aligned spins;
furthermore, this approach does not reproduce the full leading-PN result ---
which in the non-spinning case may be derived from Einstein's quadrupole formula.
To overcome both of these limitations, and following the suggestion of \Rcite{Saketh:2021sri},
we believe that in the future it will be more profitable to focus on reconstructing
the (gauge-dependent) instantaneous momentum and angular momentum fluxes.
In this case, a complete 4PM result would suffice to reconstruct the leading-PN form of the radiated energy.
Alternatively, we hope that improved unbound-to-bound mappings for spinning bodies
will further alleviate these issues.

A natural continuation of this work will therefore be to progress upwards in the perturbative series ---
both to higher PM orders and higher spin orders.
While higher PM orders will present a challenge for the integration step,
there has recently been a promising development in this area:
the first complete analytic result for the 4PM momentum impulse
including radiation-reaction effects~\cite{Dlapa:2022lmu},
wherein loop integrals with retarded propagators were also used.
With a similar basis of master integrals, it will be possible to tackle spin effects at 4PM order.
While higher PM orders present a challenge regarding the integration steps,
higher spin orders rather challenge the construction of the integrand.
In this case, it will be necessary to upgrade the spinning $\cN=2$ supersymmetric worldline
action to include more supersymmetry --- a tantalizing prospect that we leave for future work.

\sec{Acknowledgments}
We thank Alessandra Buonanno, Gregor K\"alin, Jan Plefka, Muddu Saketh, Benjamin Sauer, Chia-Hsien Shen and Justin Vines for enlightening discussions.
We are especially grateful to Jan Steinhoff for his assistance in comparing with the 4PN quadratic-in-spin Hamiltonian,
and to Mohammed Khalil for helping us produce the numerical binding energy plot.
We also thank the organizers of the ``High-Precision Gravitational Waves'' program
at the Kavli Institute for Theoretical Physics (KITP) for their hospitality.
This work is funded by the Deutsche Forschungsgemeinschaft
(DFG, German Research Foundation)
Projekt-nummer 417533893/GRK2575 ``Rethinking Quantum Field Theory''.
It was also supported in part by the National Science Foundation under Grant No.~NSF PHY-1748958.
\newpage
\appendix
\begin{widetext}
  \section{Canonical expansion of scattering angle}\label{sec:angleCan}
  Here we discuss the relationship between the canonical and covariant expansions of the conservative scattering angle $\tc$.
  As the angle is independent of the choice of SSC, only its expansion coefficients change when we expand in
  canonical rather than covariant variables.
  The canonical expansion is defined analogously to the covariant expansion in \Eqn{eq:angleCov2}:
  \begin{align}
    &
    \tc
    =
    \sum_{n=1}^3
    \Big(
    \frac{G M}{|b_{\rm can}|}
    \Big)^n
    \bigg[
      \thiic{n;0}
      -
      \sum_i \thiic{n;1,i} \frac{\hat L_{\rm can}\cdot a_{i,\rm can}}{|b_{\rm can}|}
      \\
      &\qquad\qquad
      +
      \sum_{i,j}
      \frac{a_{i,\rm can}^\mu a_{j,\rm can}^\nu}{|b_{\rm can}|^2}
      \bigg(
      -\thiic{n;2,1,i,j} \eta^{\mu\nu}
      +\thiic{n;2,2,i,j} \hat b_{\rm can}^\mu \hat b_{\rm can}^\nu
      +\thiic{n;2,3,i,j} \hat p^\mu \hat p^\nu
      +\thiic{n;2,4,i,j} \hat b_{\rm can}^\mu \hat p^\nu
      \bigg)
      \bigg]
    +\cO(S^3,G^4)\,.
    \nn
  \end{align}
  Using the definitions of the canonical variables from Sec.~\ref{sec:canonical}
  we relate the canonical expansion coefficients, $\thiic{n;A}$, to the covariant ones, $\thiio{n;A}$:
  \begin{subequations}
    \begin{align}
      \thiic{n;0}
      &=
      \Gamma\,
      \thiio{n;0}\,,
      \\
      \thiic{n;1,i}
      &=
      \Gamma\,
      \Big(
      \thiio{n;1,i}
      +
      n
      \frac{\pin}{E_i+m_i}
      \thiio{n;0}
      \Big)\,,
      \\
      \thiic{n;2,1,i,j}
      &=
      \Gamma\,
      \Big(
      \thiio{n;2,1,i,j}
      +
      \frac{(n+1)\pin}{2(E_j+m_j)}
      \thiio{n;1,i}
      +
      \frac{(n+1)\pin}{2(E_i+m_i)}
      \thiio{n;1,j}
      +
      \frac{n(n+1)\pin^2}{
        2(E_i+m_i)(E_j+m_j)
      }
      \thiio{n;0}
      \Big)\,,
      \\
      \thiic{n;2,2,i,j}
      &=
      \Gamma\,
      \Big(
      \thiio{n;2,2,i,j}
      -
      \frac{(n+2)\pin}{2(E_j+m_j)}
      \thiio{n;1,i}
      -
      \frac{(n+2)\pin}{2(E_i+m_i)}
      \thiio{n;1,j}
      -
      \frac{n(n+2)\pin^2}{2(E_i+m_i)(E_j+m_j)}
      \thiio{n;0}
      \Big)\,,
      \\
      \thiic{n;2,3,i,j}
      &=
      \Gamma\,
      \Big(
      (-1)^{i+j} \frac{\pin^2 \Gamma^2}{\mu^2}\thiio{n;2,3,i,j}
      +
      \frac{1-(-1)^{i+j}}{2}
      \frac{E_1E_2-m_1m_2+\pin^2}{m_1m_2}
      \thiio{n;2,1,1,2}
      \nn
      \\
      &\qquad\qquad
      -
      \frac{(n+1)\pin}{2(E_j+m_j)}
      \thiio{n;1,i}
      -
      \frac{(n+1)\pin}{2(E_i+m_i)}
      \thiio{n;1,j}
      -
      \frac{n(n+1)\pin^2}{2(E_i+m_i)(E_j+m_j)}
      \thiio{n;0}
      \Big)\,.
    \end{align}
  \end{subequations}
  The covariant coefficients $\thiio{n;A}$ are given in Appendix~\ref{sec:angles}.

  \section{Scattering Angle}
  \label{sec:angles}
  \allowdisplaybreaks
  In this section we print the 3PM contributions to the covariant expansion coefficients
  of the scattering angle $\theta$ defined in Eqs.~\eqref{eq:genericAngles} and~\eqref{eq:angleCov2}.
  First, we print the conservative contributions which are used for the Hamiltonian.
  Then, we print the radiative contributions.
  The coefficients that we have not printed here may be obtained by exchanging the two particles.
  The conservative non-spinning and spin-orbit coefficients are:
  \begin{align}
    &
    \thiio{3;0}
    =
    \frac{2 \left(64 \gamma
      ^6-120 \gamma ^4+60 \gamma ^2-5\right) \Gamma ^2}{3 \left(\gamma
      ^2-1\right)^3}
    -\frac{8 \gamma  \left(14 \gamma ^2+25\right) \nu }{3 \left(\gamma
      ^2-1\right)}
    -\frac{8 \left(4 \gamma ^4-12 \gamma ^2-3\right) \nu  \ \text{arccosh}\gamma
    }{\left(\gamma ^2-1\right)^{3/2}}\,,
    \\
    &
    \thiio{3;1,1}=
    -\frac{2 \gamma 
      \left(16 \gamma ^4-20 \gamma ^2+5\right) \left(5 \Gamma ^2-\delta
      \right)}{\left(\gamma ^2-1\right)^{5/2}}
    +\frac{4 \left(44 \gamma ^4+100
      \gamma ^2+41\right) \nu }{\left(\gamma ^2-1\right)^{3/2}}
    +\frac{48 \gamma  \left(\gamma ^2-6\right) \left(2 \gamma ^2+1\right) \nu 
      \ \text{arccosh}\gamma}{\left(\gamma ^2-1\right)^2}\,.
  \end{align}
  These agree with the ones in Ref.~\cite{Jakobsen:2022fcj} as expected from the discussion beneath Eq.~\eqref{eq:genericAngles}.
  The coefficients $\thiio{3;2,1,i,j}$ are:
  {\small
    \begin{align}
      &
      \thiio{3;2,1,1,1}=
      \Gamma ^2 \left(\frac{4 \left(96
        \gamma ^6-160 \gamma ^4+70 \gamma ^2-5\right)}{\left(\gamma
        ^2-1\right)^3}
      -\frac{4 \left(1772 \gamma ^6-2946 \gamma ^4+1346 \gamma
        ^2-137\right) C_{{\rm E},1}}{35 \left(\gamma ^2-1\right)^3}\right)
      \\
      &\qquad\qquad
      +\delta  \left(
      -\frac{8 \left(4 \gamma ^2-2 \gamma -1\right) \left(4
        \gamma ^2+2 \gamma -1\right)}{\left(\gamma ^2-1\right)^2}
      +
      \frac{8 \left(214 \gamma
        ^4-223 \gamma ^2+44\right) C_{{\rm E},1}}{35 \left(\gamma
        ^2-1\right)^2}
      \right)
      \nn
      \\
      &\qquad\qquad
      -\frac{16 \gamma  \left(148 \gamma ^4+374 \gamma ^2+383\right) \nu }{5
        \left(\gamma ^2-1\right)^2}
      +\frac{8
        \gamma  \left(3244 \gamma ^4+7972 \gamma ^2+4639\right) C_{{\rm E},1} \nu
      }{105 \left(\gamma ^2-1\right)^2}
      \nn
      \\
      &\qquad\qquad
      +\text{arccosh}\gamma
      \left(
      -\frac{192
        \left(\gamma ^6-8 \gamma ^4-7 \gamma ^2-1\right) \nu }{\left(\gamma
        ^2-1\right)^{5/2}}
      +
      \frac{16 \left(8 \gamma ^6-56 \gamma ^4-24 \gamma
        ^2-3\right) C_{{\rm E},1} \nu }{\left(\gamma ^2-1\right)^{5/2}}
      \right)
      \nn
      \\
      &\thiio{3;2,1,1,2}=
      \frac{4 \left(96 \gamma ^6-160 \gamma
        ^4+70 \gamma ^2-5\right) \Gamma ^2}{\left(\gamma ^2-1\right)^3}
      -\frac{32 \gamma  \left(15 \gamma ^4+46 \gamma ^2+47\right) \nu
      }{\left(\gamma ^2-1\right)^2}
      -\frac{48
        \left(4 \gamma ^6-36 \gamma ^4-35 \gamma ^2-5\right) \nu \, \text{arccosh}\gamma
      }{\left(\gamma ^2-1\right)^{5/2}}
  \end{align}}
  The coefficients $\thiio{3;2,2,i,j}$ are:
  {\small\begin{align}
    &\thiio{3;2,2,1,1}=\frac{4 \gamma  \left(9000 \gamma ^{10}+4404 \gamma
      ^8-2152 \gamma ^6-12152 \gamma ^4+8379 \gamma
      ^2-1479\right) \nu }{15 \left(\gamma ^2-1\right)^3
      \left(2 \gamma ^2-1\right)^2}
    \\&\qquad
    +\pi ^2
    \Bigg(-\frac{\left(5 \gamma ^2-3\right) \left(10
      \gamma ^4-5 \gamma ^2+9\right) \gamma ^2 \delta
    }{128 \left(\gamma ^2-1\right)^2 \left(2 \gamma
      ^2-1\right)^2}+\frac{\left(100 \gamma ^8-160 \gamma
      ^6+193 \gamma ^4-78 \gamma ^2+45\right) \gamma ^2
      \Gamma ^2}{128 \left(\gamma ^2-1\right)^2 \left(2
      \gamma ^2-1\right)^3}
    \nn   \\&\qquad
    -\frac{\left(100 \gamma ^9-160
      \gamma ^7-60 \gamma ^6+193 \gamma ^5-12 \gamma
      ^4-78 \gamma ^3+12 \gamma ^2+45 \gamma -36\right)
      \gamma ^2 \nu }{64 \left(\gamma ^2-1\right)^2
      \left(2 \gamma ^2-1\right)^3}\Bigg)
    +\delta 
    \Bigg(\frac{10 \left(4 \gamma ^2-2 \gamma -1\right)
      \left(4 \gamma ^2+2 \gamma -1\right)}{\left(\gamma
      ^2-1\right)^2}
    \nn   \\&\qquad
    -\frac{2 \left(1192 \gamma ^4-1382
      \gamma ^2+295\right) C_{{\rm E},1}}{35 \left(\gamma
      ^2-1\right)^2}\Bigg)
    -\frac{4 \gamma  \left(10744
      \gamma ^6+13474 \gamma ^4+2665 \gamma
      ^2+9237\right) C_{{\rm E},1} \nu }{105 \left(\gamma
      ^2-1\right)^3}
    \nn   \\&\qquad
    +\Gamma ^2 \left(\frac{2 \left(6568
      \gamma ^6-11114 \gamma ^4+5079 \gamma ^2-463\right)
      C_{{\rm E},1}}{35 \left(\gamma
      ^2-1\right)^3}-\frac{2 \left(960 \gamma ^{10}-2560
      \gamma ^8+2540 \gamma ^6-1150 \gamma ^4+225 \gamma
      ^2-13\right)}{\left(\gamma ^2-1\right)^3 \left(2
      \gamma ^2-1\right)^2}\right)
    \nn   \\&\qquad
    +\text{arccosh}\gamma
    \left(\frac{16 \gamma ^2 \left(60 \gamma ^{10}-600
      \gamma ^8+551 \gamma ^6-63 \gamma ^4-63 \gamma
      ^2+15\right) \nu }{\left(\gamma ^2-1\right)^{7/2}
      \left(2 \gamma ^2-1\right)^2}-\frac{32 \left(8
      \gamma ^8-56 \gamma ^6+26 \gamma ^4-18 \gamma
      ^2-3\right) C_{{\rm E},1} \nu }{\left(\gamma
      ^2-1\right)^{7/2}}\right)
    \nn   \\&
    \thiio{3;2,2,1,2}
    =-\frac{2 \left(960 \gamma ^{10}-2560 \gamma ^8+2540
      \gamma ^6-1150 \gamma ^4+225 \gamma ^2-13\right)
      \Gamma ^2}{\left(\gamma ^2-1\right)^3 \left(2
      \gamma ^2-1\right)^2}
    \\&\qquad
    +\frac{4 \gamma  \left(1800
      \gamma ^{10}+2020 \gamma ^8-424 \gamma ^6-3032
      \gamma ^4+1703 \gamma ^2-231\right) \nu }{3
      \left(\gamma ^2-1\right)^3 \left(2 \gamma
      ^2-1\right)^2}
    \nn\\&\qquad
    +\frac{16 \left(60 \gamma ^{12}-664
      \gamma ^{10}+519 \gamma ^8-31 \gamma ^6-43 \gamma
      ^4+7 \gamma ^2-1\right) \nu\  \text{arccosh}\gamma}{\left(\gamma ^2-1\right)^{7/2} \left(2 \gamma
      ^2-1\right)^2}
    \nn\\&\qquad
    +\pi ^2 \left(\frac{3 \left(\gamma
      ^2+1\right) \left(5 \gamma ^4-4 \gamma ^2+3\right)
      \gamma ^2 \Gamma ^2}{32 \left(\gamma ^2-1\right)^2
      \left(2 \gamma ^2-1\right)^3}+\frac{\left(100
      \gamma ^8-60 \gamma ^7-160 \gamma ^6-12 \gamma
      ^5+193 \gamma ^4+12 \gamma ^3-78 \gamma ^2-36
      \gamma +45\right) \gamma ^2 \nu }{64 \left(\gamma
      ^2-1\right)^2 \left(2 \gamma ^2-1\right)^3}\right)
    \nn
  \end{align}}
  The coefficients $\thiio{3;2,3,i,j}$ are:
  {\small\begin{align}
    &\thiio{3;2,3,1,1}=\frac{\gamma  \left(18624 \gamma ^8+24848 \gamma
      ^6-45192 \gamma ^4-58631 \gamma ^2+36351\right) \nu
    }{15 \left(\gamma ^2-1\right)^4 \left(2 \gamma
      ^2-1\right)}
    +\delta  \Bigg(\frac{704 \gamma ^6-880
      \gamma ^4+312 \gamma ^2-23}{2 \left(\gamma
      ^2-1\right)^3 \left(2 \gamma ^2-1\right)}
    \\&\qquad
    -\frac{2
      \left(1376 \gamma ^4-1294 \gamma ^2+233\right)
      C_{{\rm E},1}}{35 \left(\gamma
      ^2-1\right)^3}\Bigg)-\frac{4 \gamma  \left(8720
      \gamma ^6+14894 \gamma ^4-22663 \gamma
      ^2-37071\right) C_{{\rm E},1} \nu }{105 \left(\gamma
      ^2-1\right)^4}
    \nn\\&\qquad
    +\Gamma ^2 \left(\frac{2 \left(4064
      \gamma ^6-6562 \gamma ^4+2997 \gamma ^2-359\right)
      C_{{\rm E},1}}{35 \left(\gamma
      ^2-1\right)^4}-\frac{1728 \gamma ^8-3664 \gamma
      ^6+2584 \gamma ^4-673 \gamma ^2+41}{2 \left(\gamma
      ^2-1\right)^4 \left(2 \gamma
      ^2-1\right)}\right)
    \nn\\&\qquad
    +\text{arccosh}\gamma
    \left(\frac{64 \left(3 \gamma ^8-35 \gamma ^6+9
      \gamma ^4+42 \gamma ^2+6\right) \nu }{\left(\gamma
      ^2-1\right)^{9/2}}-\frac{16 \left(8 \gamma ^8-80
      \gamma ^6+44 \gamma ^4+99 \gamma ^2+15\right)
      C_{{\rm E},1} \nu }{\left(\gamma
      ^2-1\right)^{9/2}}\right)
    \nn\\&
    \thiio{3;2,3,1,2}=\frac{4 \gamma  \left(96 \gamma ^6-148 \gamma ^4+55
      \gamma ^2-1\right) \Gamma ^2}{\left(\gamma
      ^2-1\right)^4}-\frac{16 \gamma  \left(12 \gamma
      ^8-136 \gamma ^6-21 \gamma ^4+210 \gamma
      ^2+88\right) \nu\ \text{arccosh}\gamma}{\left(\gamma ^2-1\right)^{9/2}}
    \\&\qquad
    -\frac{\left(2880
      \gamma ^{10}+7712 \gamma ^8-5664 \gamma ^6-22688
      \gamma ^4+9219 \gamma ^2+1197\right) \nu }{3
      \left(\gamma ^2-1\right)^4 \left(2 \gamma
      ^2-1\right)}
    \nn
  \end{align}}
  The coefficients $\thiio{3;2,4,i,j}$ are:
  {\small\begin{align}
    &\thiio{3;2,4,1,1}=\pi  \Bigg(-\frac{\left(80 \gamma ^9-144 \gamma ^7-12
      \gamma ^6+42 \gamma ^5+33 \gamma ^4+32 \gamma ^3-21
      \gamma ^2-18 \gamma +6\right) \nu }{4 \left(\gamma
      ^2-1\right)^{5/2} \left(2 \gamma
      ^2-1\right)^2}-\frac{3 \left(30 \gamma ^3-15 \gamma
      ^2-6 \gamma +1\right) C_{{\rm E},1} \nu }{4
      \left(\gamma ^2-1\right)^{3/2}}
    \\&\qquad
    +\Gamma ^2
    \left(\frac{80 \gamma ^8-104 \gamma ^6+12 \gamma
      ^5+20 \gamma ^4-15 \gamma ^3+18 \gamma ^2+9 \gamma
      -6}{8 \left(\gamma ^2-1\right)^{5/2} \left(2 \gamma
      ^2-1\right)^2}+\frac{3 \left(30 \gamma ^4+15 \gamma
      ^3-21 \gamma ^2-\gamma +3\right) C_{{\rm E},1}}{8
      \left(\gamma ^2-1\right)^{5/2}}\right)
    \nn\\&\qquad
    +\delta 
    \left(-\frac{80 \gamma ^8-104 \gamma ^6-12 \gamma
      ^5+20 \gamma ^4+15 \gamma ^3+18 \gamma ^2-9 \gamma
      -6}{8 \left(\gamma ^2-1\right)^{5/2} \left(2 \gamma
      ^2-1\right)^2}-\frac{3 \left(30 \gamma ^4-15 \gamma
      ^3-21 \gamma ^2+\gamma +3\right) C_{{\rm E},1}}{8
      \left(\gamma ^2-1\right)^{5/2}}\right)\Bigg)
    \nn\\
    &\thiio{3;2,4,1,2}=\pi  \Bigg(-\frac{\left(20 \gamma ^7+12 \gamma ^6-21
      \gamma ^5-24 \gamma ^4+4 \gamma ^3+15 \gamma ^2+3
      \gamma -3\right) \Gamma ^2}{4 \left(\gamma
      ^2-1\right)^{5/2} \left(2 \gamma
      ^2-1\right)^2}
    \\&\qquad
    +\frac{\left(20 \gamma ^7-12 \gamma
      ^6-21 \gamma ^5+24 \gamma ^4+4 \gamma ^3-15 \gamma
      ^2+3 \gamma +3\right) \delta }{4 \left(\gamma
      ^2-1\right)^{5/2} \left(2 \gamma
      ^2-1\right)^2}
    \nn\\&\qquad
    +\frac{\left(120 \gamma ^8-56 \gamma
      ^7-194 \gamma ^6+48 \gamma ^5+124 \gamma ^4-\gamma
      ^3-36 \gamma ^2-9 \gamma +6\right) \nu }{4
      \left(\gamma ^2-1\right)^{5/2} \left(2 \gamma
      ^2-1\right)^2}\Bigg)
    \nn
  \end{align}}
  Let us then print the radiative contributions.
  All coefficients are proportional to the universal function $\mathcal{I}(v)$ except $\theta^{3;2,4,i,j}_{k,\rm rad}$.
  These coefficients depend on the particle label $k$.
  {\small\begin{align}
    \thiir{3;0}&=
    \frac{4 \left(1-2 \gamma ^2\right)^2 \nu
    }{\left(\gamma ^2-1\right)^{3/2}}
    \mathcal{I}(v)
    \\
    \thiir{3;1,1}&=
    -\frac{24 \gamma  \left(2 \gamma ^2-1\right) \nu
    }{\gamma ^2-1}
    \mathcal{I}(v)
    \\
    \thiir{3;2,1,1,1}&=
    -\frac{16 \nu  \left(\gamma ^4 (4
      C_{{\rm E},1}-6)+\gamma ^2 (6-4
      C_{{\rm E},1})+C_{{\rm E},1}-1\right)}{\left(\gamma
      ^2-1\right)^{3/2}}
    \mathcal{I}(v)
    \\
    \thiir{3;2,1,1,2}&=
    \frac{16 \left(6 \gamma ^4-6 \gamma ^2+1\right) \nu
    }{\left(\gamma ^2-1\right)^{3/2}}
    \mathcal{I}(v)
    \\
    \thiir{3;2,2,1,1}&=
    \frac{8 \nu  \left(\gamma ^4 (16
      C_{{\rm E},1}-15)+\gamma ^2 (15-16 C_{{\rm E},1})+4
      (C_{{\rm E},1}-1)\right)}{\left(\gamma
      ^2-1\right)^{3/2}}
    \mathcal{I}(v)
    \\
    \thiir{3;2,2,1,2}&=
    -\frac{8 \left(15 \gamma ^4-15 \gamma ^2+4\right) \nu
    }{\left(\gamma ^2-1\right)^{3/2}}
    \mathcal{I}(v)
    \\
    \thiir{3;2,3,1,1}&=
    \frac{16 \nu  \left(\gamma ^4 (4
      C_{{\rm E},1}-6)+\gamma ^2 (6-4
      C_{{\rm E},1})+C_{{\rm E},1}-1\right)}{\left(\gamma
      ^2-1\right)^{5/2}}
    \mathcal{I}(v)
    \\
    \thiir{3;2,3,1,2}&=
    \frac{16 \gamma  \left(6 \gamma ^4-6 \gamma
      ^2+1\right) \nu }{\left(\gamma ^2-1\right)^{5/2}}
    \mathcal{I}(v)
  \end{align}}
  The coefficients $\theta^{3;2,4,i,j}_{k,\rm rad}$ are:
  {\small\begin{align}
    &\theta^{(3;2,4,1,1)}_{1,\rm rad}=
    \frac{3 \pi  \gamma  \left(28 \gamma ^8-80 \gamma
      ^6+41 \gamma ^4+34 \gamma ^2-15\right) \nu\ \text{arccosh}\gamma}{32 \left(\gamma ^2-1\right)^{7/2}
      \left(2 \gamma ^2-1\right)}
    \\&\qquad
    -\frac{3 \pi  \left(14
      \gamma ^7+14 \gamma ^6+101 \gamma ^5-323 \gamma
      ^4+236 \gamma ^3-60 \gamma ^2-23 \gamma +25\right)
      \nu  \log \left(\frac{\gamma +1}{2}\right)}{16
      (\gamma -1)^2 (\gamma +1)^3 \left(2 \gamma
      ^2-1\right)}
    \nn\\&\qquad
    +\frac{\pi  \left(280 \gamma ^{10}+470
      \gamma ^9+1930 \gamma ^8+5397 \gamma ^7-88373
      \gamma ^6+244865 \gamma ^5-273845 \gamma ^4+73199
      \gamma ^3+112249 \gamma ^2-102411 \gamma
      +25759\right) \nu }{160 (\gamma -1)^3 (\gamma +1)^5
      \left(2 \gamma ^2-1\right)}
    \nn\\&\qquad
    +C_{{\rm E},1}
    \Bigg(\frac{3 \pi  \gamma  \left(98 \gamma ^6-239
      \gamma ^4+164 \gamma ^2-39\right) \nu \ \text{arccosh}\gamma}{64 \left(\gamma
      ^2-1\right)^{7/2}}-\frac{3 \pi  \left(49 \gamma
      ^5+169 \gamma ^4-306 \gamma ^3+150 \gamma ^2-63
      \gamma +33\right) \nu  \log \left(\frac{\gamma
        +1}{2}\right)}{32 (\gamma -1)^2 (\gamma
      +1)^3}
    \nn\\&\qquad
    -\frac{\pi  \left(1575 \gamma ^9-810 \gamma
      ^8+1020 \gamma ^7+4140 \gamma ^6+26442 \gamma
      ^5-196568 \gamma ^4+442476 \gamma ^3-521244 \gamma
      ^2+321447 \gamma -80398\right) \nu }{640 (\gamma
      -1)^3 (\gamma +1)^5}\Bigg)
    \nn
    \\
    &\theta^{(3;2,4,2,2)}_{1,\rm rad}=
    -\frac{3 \pi  \left(2 \gamma ^4-13 \gamma ^2+15\right)
      \gamma ^2 \nu\  \text{arccosh}\gamma}{16
      \left(\gamma ^2-1\right)^{7/2} \left(2 \gamma
      ^2-1\right)}+\frac{3 \pi  \left(-110 \gamma ^5+259
      \gamma ^4-186 \gamma ^3+8 \gamma ^2+40 \gamma
      -19\right) \nu  \log \left(\frac{\gamma
        +1}{2}\right)}{8 (\gamma -1)^2 (\gamma +1)^3
      \left(2 \gamma ^2-1\right)}
    \\&\qquad
    +\frac{\pi  \left(-2070
      \gamma ^9+2740 \gamma ^8+20777 \gamma ^7-153646
      \gamma ^6+408765 \gamma ^5-463000 \gamma ^4+119039
      \gamma ^3+196158 \gamma ^2-168991 \gamma
      +40708\right) \nu }{160 (\gamma -1)^3 (\gamma +1)^5
      \left(2 \gamma ^2-1\right)}
    \nn\\&\qquad
    +C_{{\rm E},2}
    \Bigg(-\frac{3 \pi  \left(30 \gamma ^4-59 \gamma
      ^2+21\right) \gamma ^2 \nu \ \text{arccosh}\gamma}{32 \left(\gamma ^2-1\right)^{7/2}}-\frac{3 \pi 
      \left(65 \gamma ^4-190 \gamma ^3+156 \gamma ^2-74
      \gamma +27\right) \nu  \log \left(\frac{\gamma
        +1}{2}\right)}{16 (\gamma -1)^2 (\gamma
      +1)^3}
    \nn\\&\qquad
    +\frac{\pi  \left(-1075 \gamma ^7+980 \gamma
      ^6-6738 \gamma ^5+73384 \gamma ^4-198749 \gamma
      ^3+236162 \gamma ^2-135438 \gamma +30994\right) \nu
    }{160 (\gamma -1)^3 (\gamma +1)^5}\Bigg)
    \nn
    \\
    &\theta^{(3;2,4,1,2)}_{1,\rm rad}=
    -\frac{3 \pi  \left(8 \gamma ^6-22 \gamma ^4+9 \gamma
      ^2+9\right) \gamma ^2 \nu \ \text{arccosh}\gamma}{16
      \left(\gamma ^2-1\right)^{7/2} \left(2 \gamma
      ^2-1\right)}+\frac{3 \pi  \left(17 \gamma ^6-63
      \gamma ^5+107 \gamma ^4-81 \gamma ^3+8 \gamma ^2+16
      \gamma -8\right) \nu  \log \left(\frac{\gamma
        +1}{2}\right)}{4 (\gamma -1)^2 (\gamma +1)^3
      \left(2 \gamma ^2-1\right)}
    \nn\\&\qquad
    +\frac{\pi  \left(-630
      \gamma ^{10}-480 \gamma ^9+4481 \gamma ^8+11476
      \gamma ^7-88357 \gamma ^6+195970 \gamma ^5-191553
      \gamma ^4+34776 \gamma ^3+87891 \gamma ^2-69710
      \gamma +16328\right) \nu }{64 (\gamma -1)^3 (\gamma
      +1)^5 \left(2 \gamma ^2-1\right)}
    \\
    &\theta^{(3;2,4,2,1)}_{1,\rm rad}=
    \frac{3 \pi  \gamma  \left(4 \gamma ^6-24 \gamma ^4+33
      \gamma ^2-9\right) \nu \ \text{arccosh}\gamma}{16
      \left(\gamma ^2-1\right)^{7/2} \left(2 \gamma
      ^2-1\right)}+\frac{3 \pi  \left(20 \gamma ^6-76
      \gamma ^5+120 \gamma ^4-73 \gamma ^3-\gamma ^2+25
      \gamma -11\right) \nu  \log \left(\frac{\gamma
        +1}{2}\right)}{4 (\gamma -1)^2 (\gamma +1)^3
      \left(2 \gamma ^2-1\right)}
    \nn\\&\qquad
    +\frac{\pi  \left(-98
      \gamma ^9+1488 \gamma ^8+9431 \gamma ^7-75972
      \gamma ^6+188037 \gamma ^5-200914 \gamma ^4+48417
      \gamma ^3+85464 \gamma ^2-73947 \gamma
      +17902\right) \nu }{64 (\gamma -1)^3 (\gamma +1)^5
      \left(2 \gamma ^2-1\right)}
  \end{align}}

  \section{Hamiltonian coefficients}\label{sec:HamCoe}

  Here we give the Hamiltonian coefficients in terms of those of the scattering angle.
  We do not print the expressions for the spin-spin coefficients at 3PM order which are found in the ancillary file.
  We start with the simpler spinless and spin-orbit coefficients.
  At 1PM order we find:
  \begin{align}
    \cii{1;0}(\vct{p}^2)
    &=
    \left.
    -\frac{\pin^2}{2 E \xi}
    \thiic{1;0}
    \right|_{\pin\to|\mathbf{p}|}
    \ ,
    \\
    \cii{1;1,i}(\vct{p}^2)
    &=
    \left.
    -\frac{\pin}{2 E\xi}
    \thiic{1;1,i}
    \right|_{\pin\to|\mathbf{p}|}
    \ .
    \nn
  \end{align}
  At 2PM order we find:
  \begin{align}
    \cii{2;0}(\vct{p}^2)
    &=\left.
    -\frac{\pin^2}{\pi E\xi}
    \thiic{2;0}
    +
    \frac{1}{8\pin}
    \cD\Big[
      \frac{\pin^3}{E\xi}
      \big(
      \thiic{1;0}
      \big)^2
      \Big]
    \right|_{\pin\to|\mathbf{p}|}
    \ ,
    \\
    \cii{2;1,i}(\vct{p}^2)
    &=\left.
    -\frac{\pin}{\pi E\xi}
    \thiic{2;1,i}
    +
    \frac{1}{4\pin^3}
    \cD\Big[
      \frac{\pin^4}{E\xi}
      \thiic{1;0}\thiic{1;1,i}
      \Big]
    \right|_{\pin\to|\mathbf{p}|}
    \ .
    \nn
  \end{align}
  Finally we reprint the non-spinning and spin-orbit coefficients at 3PM order from Eq.~\eqref{eq:HamCoe}:
  \begin{align}
    \cii{3;0}(\mathbf{p}^2)
    &=
    \left.
    -\frac{\pin^2}{4E\xi}
    \thiic{3;0}
    +
    \frac1{2\pi\pin^2}
    \cD\Big[
      \frac{\pin^4}{E\xi}
      \thiic{1;0}
      \thiic{2;0}
      \Big]
    -
    \frac1{48\pin^2}
    \cD^2\Big[
      \frac{\pin^4}{E\xi}
      (\thii{1;0})^3
      \Big]
    \right|_{\pin\to|\mathbf{p}|}
    \ ,
    \\
    \cii{3;1,i}(\mathbf{p}^2)
    &=\left.
    -\frac{\pin}{4E\xi}
    \thiic{3;1,i}
    +
    \frac1{2\pi\pin^4}
    \cD\Big[\frac{\pin^5}{E\xi}
      \big(
      \thiic{1;0}
      \thiic{2;1,i}
      +
      \thiic{2;0}
      \thiic{1;1,i}
      \big)
      \Big]
    -
    \frac1{16\pin^4}
    \cD^2\Big[
      \frac{\pin^5}{E\xi}
      (\thiic{1;0})^2
      \thiic{1;1,i}
      \Big]
    \right|_{\pin\to|\mathbf{p}|}
    \ .\nn
  \end{align}
  The spin-spin coefficients do not seem to obey the same simplicity as the above spinless and spin-orbit coefficients.
  At the first Post-Minkowskian order we find:
  \begin{subequations}
    \begin{align}
      \cii{1;2,1,i,j}(\vct{p}^2)
      &=
      \left.
      -\frac{\pin^2}{4E\xi}
      \thiic{1;2,1,i,j}
      \right|_{\pin\to|\mathbf{p}|}
      \ ,
      \\
      \cii{1;2,2,i,j}(\vct{p}^2)
      &=
      \left.
      -\frac{3\pin^2}{8E\xi}
      \thiic{1;2,2,i,j}
      +
      \frac{3\pin^2}{16E\xi}
      \frac{
        \thiic{1;1,i}\thiic{1;1,j}
      }{
        \thiic{1;0}
      }
      \right|_{\pin\to|\mathbf{p}|}
      \ ,
      \\
      \cii{1;2,3,i,j}(\vct{p}^2)
      &=
      \left.
      -\frac{1}{4E\xi}
      \big(
      \thiic{1;2,3,i,j}
      -
      \frac12 \thiic{1;2,2,i,j}
      \big)
      -
      \frac{1}{16E\xi}
      \frac{
        \thiic{1;1,i}\thiic{1;1,j}
      }{
        \thiic{1;0}
      }
      \right|_{\pin\to|\mathbf{p}|}
      \ .
    \end{align}
  \end{subequations}
  At second Post-Minkowskian order we find:
  \begin{subequations}
    \begin{align}
      \cii{2;2,1,i,j}(\vct{p}^2)
      &=\left.
      -\frac{2\pin^2}{3\pi E\xi}
      \thiic{2;2,1,i,j}
      +
      \frac{1}{32\pin^3}
      \cD\Big[
        \frac{\pin^5}{E\xi}
        \Big(
        3\thiic{1;1,i}\thiic{1;1,j}
        +
        4\thiic{1;0}\thiic{1;2,1,i,j}
        \Big)
        \Big]
      \right|_{\pin\to|\mathbf{p}|}
      \\
      \cii{2;2,2,i,j}(\vct{p}^2)
      &=
      -\frac{\pin^2}{E\xi}
      \bigg(
      \frac{8\thiic{2;2,2,i,j}}{9\pi}
      +\frac{\thiic{1;0}\thiic{1;2,2,i,j}}{8}
      -\frac{\thiic{1;1,i}\thiic{1;1,j}}{16}
      +
      \frac{\pin}{16}
      \Big(
      \frac{\thiic{1;1,i}}{m_i}
      +
      \frac{\thiic{1;2,j}}{m_j}
      \Big)
      \thiic{1;2,2,i,j}
      \bigg)
      \\
      &\qquad
      +
      \frac{1}{32\pin^3}
      \cD\Big[
        \frac{\pin^5}{E\xi}
        \Big(
        6\thiic{1;0}\thiic{1;2,2,i,j}
        -
        7\thiic{1;1,i}\thiic{1;1,j}
        \Big)
        \Big]
      -\frac{4\pin^2}{9\pi E\xi}
      \frac{
        \thiic{2;0}\thiic{1;1,i}\thiic{1;1,j}
      }{
        \big(\thiic{1;0}\big)^2
      }
      \nn
      \\
      &\qquad
      \left.
      +\frac{\pin^2}{E\xi\thiic{1;0}}
      \bigg(
      \frac{\pin}{32}
      \Big(
      \frac{\thiic{1;1,i}}{m_i}
      +
      \frac{\thiic{1;1,j}}{m_j}
      \Big)
      \thiic{1;1,i}\thiic{1;1,j}
      +\frac2{9\pi}
      \big(
      \thiic{1;1,i}\thiic{2;1,j}+
      \thiic{2;1,i}\thiic{1;1,j}
      \big)
      \bigg)
      \right|_{\pin\to|\mathbf{p}|}
      \nn
      \\
      \cii{2;2,3,i,j}(\vct{p}^2)
      &=
      -\frac{1}{E\xi}
      \bigg(
      \frac{
        2  \thiic{2;2,3,i,j}
      }{3\pi}
      -
      \frac{
        2 \thiic{2;2,2,i,j}
      }{9\pi}
      -
      \frac{
        \thiic{1;0}\thiic{1;2,2,i,j}
      }{8}
      -
      \frac{\pin}{16}
      \Big(
      \frac{\thiic{1;1,i}}{m_i}
      +
      \frac{\thiic{1;2,j}}{m_j}
      \Big)
      \thiic{1;2,2,i,j}
      \bigg)
      \\
      &\qquad
      +
      \frac{\thiic{1;1,i}\thiic{1;1,j}}{16}
      -
      \frac{1}{32\pin^5}
      \cD\Big[
        \frac{\pin^5}{E\xi}
        \Big(
        2\thiic{1;0}\thiic{1;2,2,i,j}
        +
        \thiic{1;1,i}\thiic{1;1,j}
        -
        4\thiic{1;0}\thiic{1;2,3,i,j}
        \Big)
        \Big]
      \nn
      \\
      &\qquad
      -\frac{1}{E\xi\thiic{1;0}}
      \bigg(
      \frac{\pin}{32}
      \Big(
      \frac{\thiic{1;1,i}}{m_i}
      +
      \frac{\thiic{1;1,j}}{m_j}
      \Big)
      \thiic{1;1,i}\thiic{1;1,j}
      +\frac1{18\pi}
      \big(
      \thiic{1;1,i}\thiic{2;1,j}+
      \thiic{2;1,i}\thiic{1;1,j}
      \big)
      \bigg)
      \nn
      \\
      &\qquad
      \left.
      +
      \frac{1}{9\pi E\xi}
      \frac{
        \thiic{2;0}\thiic{1;1,i}\thiic{1;1,j}
      }{
        \big(\thiic{1;0}\big)^2
      }
      \right|_{\pin\to|\mathbf{p}|}
      \nn
    \end{align}
  \end{subequations}
  The 3PM spin-spin coefficients are found in the ancillary file.

\end{widetext}
\bibliographystyle{JHEP}
\bibliography{../bib/wqft_spin}

\end{document}